\documentclass[prd,aps,12pt,showpacs,nofootinbib,tightenlines]{revtex4}
\usepackage[colorlinks,urlcolor=blue,linkcolor=blue,citecolor=blue]{hyperref} %
\usepackage{epsfig,graphicx,amsmath,amssymb,float,color,xspace}
\usepackage{multirow}
\begin{document}

\newcommand{\beq}{\begin{eqnarray}}
\newcommand{\eeq}{\end{eqnarray}}
\newcommand{\non}{\nonumber \\ }
\newcommand{\psl}{ p \hspace{-2.0truemm}/ }
\newcommand{\qsl}{ q \hspace{-2.0truemm}/ }
\newcommand{\epsl}{ \epsilon \hspace{-2.0truemm}/ }
\newcommand{\nsl}{ n \hspace{-2.2truemm}/ }
\newcommand{\vsl}{ v \hspace{-2.2truemm}/ }

\newcommand{\jpsi}{ J/\Psi }
\newcommand{\cala}{ {\cal A} }
\newcommand{\calb}{ {\cal B} }


\def \cpc{ { Chin. Phys. C } }
\def \ctp{ { Commun. Theor. Phys. } }
\def \csb{{ Chin. Sci. Bull. } }
\def \sbu{{ Sci. Bull.  } }

\def \epjc{{ Eur. Phys. J. C} }
\def \ijmpa{ { Int. J. Mod. Phys. A } }
\def \jhep{{ JHEP } }
\def \jpg{ { J. Phys. G} }
\def \mpla{ { Mod. Phys. Lett. A } }
\def \npb{ { Nucl. Phys. B} }
\def \plb{ { Phys. Lett. B} }
\def \ppnp{ Prog.Part. $\&$ Nucl. Phys. }
\def \pr{ { Phys. Rep.} }
\def \prd{ { Phys. Rev. D} }
\def \prl{ { Phys. Rev. Lett.}  }
\def \ptp{ { Prog. Theor. Phys. }  }
\def \zpc{ { Z. Phys. C}  }

\def \thl {{\theta_\ell}}
\def \thK {{\theta_{K}}}
\def \re{\text{Re}}
\def \im{\text{Im}}
\def \eff{{\text{eff}}}
\def\Sin{\text{sin}}
\def\Cos{\text{cos}}

\title{Study of  \texorpdfstring{$B_s\to \phi \ell^+ \ell^-$}{decay channel}  decays  in  the PQCD factorization approach with lattice QCD input}
\author{Su-Ping  Jin$^{1}$ }  \email{2223919088@qq.com}
\author{Zhen-Jun Xiao$^{1,2}$  } \email{xiaozhenjun@njnu.edu.cn; Corresponding author}
\affiliation{1.  Department of Physics and Institute of Theoretical Physics,
Nanjing Normal University, Nanjing, Jiangsu 210023, People's Republic of China,}
\affiliation{2. Jiangsu Key Laboratory for Numerical Simulation of Large Scale Complex Systems,
Nanjing Normal University, Nanjing 210023, People's Republic of China}
\date{\today}
\begin{abstract}
In this paper,  we studied systematically the semileptonic decays $B_s \to \phi l^+ l^-$ with $l^-=(e^-,\mu^-,\tau^-)$  by using the perturbative QCD (PQCD)
and the  ``PQCD+Lattice" factorization approach, respectively.
We first evaluated all relevant form factors  $F_i(q^2)$ in the low $q^2$ region using the PQCD approach, and we  also took the available lattice QCD results
at the high-$q^2$ region as additional input to improve  the extrapolation  of    $F_i(q^2)$ from the low-$q^2$ region to the endpoint $q^2_{max}$.
 We then calculated the branching ratios and many other physical observables: $A_{FB}^{l}$,  $F_L^{\phi}$,  $S_{3,4,7}$, $A_{5,6,8,9}$
 and the  clean angular observables $P_{1,2,3} $ and $P^{\prime}_{4,5,6,8}$.
From our studies, we find the following points:
(a) the PQCD and ``PQCD+Lattice" predictions of ${\cal B}( B_s \to \phi \mu^+ \mu^-)$ are about $7\times 10^{-7}$, which agree well with the
LHCb measured values  and the QCD sum rule prediction within still large errors;
(b) we defined and calculated the ratios of the branching ratios $R_\phi^{e\mu}$ and  $R_\phi^{\mu\tau}$;
(c) the PQCD and "PQCD+Lattice" predictions of the longitudinal polarization $F_L$, the CP averaged angular coefficients $S_{3,4,7} $
and  the  CP asymmetry angular coefficients $A_{5,6,8,9}$, agree with the LHCb measurements in all considered bins within the still large experimental errors;
and (d) for those currently still unknown observables  $R_\phi^{e\mu}, R_\phi^{\mu\tau}, A_{FB}^{l}, P_{1,2,3}$ and $P^{\prime}_{4,5,6,8}$,
we suggest LHCb and Belle-II Collaboration to measure them in their experiments.
\end{abstract}


\maketitle

\section{Introduction}\label{sec:1} 

In the Standard Model (SM) of  particle physics, one treats these three generations of the charged leptons $\ell^-=(e^-,\mu^-,\tau^-)$ as exact copies of each other.
These charged leptons behave in the same way but differ only in the masses determined by their Yukawa coupling to the Higgs boson.
The lepton flavor universality (LFU),   i.e. the equality of the coupling to the all electroweak gauge bosons among three families of leptons, has been regarded as an exact symmetry for quite a long time\cite{Dey:2018arx}.
In recent years, however, some  physics observables associated with the flavor-changing neutral current (FCNC) transitions $b \to s \ell \ell $ have exhibited deviations from
the SM expectations. These include the LFU-violating(LFUV) ratios $R_K$ and $R_{K^\ast} $ \cite{Aaij:2014ora,Aaij:2017vbb}, whose measurements deviates from $\mu-e$
universality  \cite{Bordone:2016gaq,Hiller:2003js,Hiller2015} by around $2.5\sigma$.
More notably, the measurements of the angular observable $P^\prime_5$  of $B \to K^\ast \mu^+\mu^-$ decay in the large recoil region
\cite{Descotes-Genon:2013wba,Descotes-Genon:2013vna,Matias:2012xw,DescotesGenon:2012zf,Matias:2014jua}
as reported by the LHCb \cite{Aaij:2013qta,Aaij:2015oid} and Belle Collaboration \cite{Abdesselam:2016llu}  point to a deviation of about $3\sigma$ with respect to the SM prediction \cite{Descotes-Genon:2014uoa}.

As is well known, the FCNC $b \to s$ transition is forbidden at tree-level, but proceeds by way of loop diagrams with a very low rate.
Due to the strong suppression within SM,  such kinds of  FCNC  decays may be sensitive to the possible new physics (NP) effects.
Therefore, the semileptonic $b \to s \ell \ell$ decay has received striking attentions by means of measurements of  the inclusive $B \to X_s\ell^+\ell^-$ and/or
the exclusive $B \to K^{(\ast)}\ell^+\ell^-$ decays and their comparison with  the SM predictions.
Besides the decay rates, many angular observables of  the semileptonic $B \to K^{\ast}\mu^+\mu^-$ decays have also been measured previously \cite{Aaij:2013qta,Aaij:2015oid,Abdesselam:2016llu}.
The precision of the experimental measurements will also be expected to upgrade remarkably  in the forthcoming year.

The semileptonic decay $B_s \to \phi \mu^+\mu^-$, which is closely relevant to the decay $B \to K^{\ast}\mu^+\mu^-$,
offers an alternative scene to check out the same fundamental quark process, in a different hadronic background.
On the theoretical side, various studies on the quark level $b\to s$ transition and the exclusive $B_{(s)} \to V \ell^+\ell^-$ decays by using  rather different theories
or models have been performed within the SM, such as the constituent quark model or  covariant quark model \cite{Deandrea:2001qs,Dubnicka:2016nyy},
the light front quark model \cite{Geng:2003su}, the QCD factorization (QCDF) \cite{Bobeth:2008ij} and the light-cone sum rule (LCSR)
\cite{Ball:2004rg,Altmannshofer:2014rta,Wu:2006rd,Straub:2015ica,Gao2020,Descotes-Genon:2015uva},
and beyond the SM, such as the universal extra dimension \cite{Mohanta:2006ae,Li:2011yn} and the supersymmetric theory \cite{Yuan-Guo:2013vpa}.
On the experimental side,  the $B_s \to \phi \mu^+\mu^-$ decay mode was first observed and studied by the CDF collaboration \cite{Aaltonen:2011cn} and subsequently by the
LHCb collaboration \cite{Aaij:2013aln,Aaij:2015esa,Aaij:2103,Aaij:2021pkz,LHCb-2107}.
Beyond the measurement of the branching ratio, a rich phenomenology of various kinematical distributions can be presented.
While the angular distributions was found to be consistent with the SM expectations obtained in Refs.~\cite{Altmannshofer:2014rta,Straub:2015ica},
however,  LHCb  also observed a deficit with respect to the SM prediction for the branching ratio $B^0_s \to \phi \mu^+ \mu^-$
in the low-$q^2$ region: the tension between the theory and experiment is  about $3\sigma$  in the region $1.0\leq q^2 \leq 6.0 GeV^2$,
where the form factors are evaluated by using the combined fit of lattice and the  LCSR results \cite{Altmannshofer:2014rta,Straub:2015ica}.

In a previous paper \cite{Jin:2020jtu}, the semileptonic $B_s \to K^{(\ast)}  \ell^+\ell^+$ decays have been
studied by us using the perturbative QCD (PQCD) factorization approach \cite{Keum:2000wi,Lu:2000em,li2003,Li:2009pr,Fan:2012kn,Fan:2013qz,Wang:2012ab,Wang:2013ix,Xiao:2011tx,Ali:2007ff,xiao18a,xiao18b}.
In this paper, we will make  systematic  studies  for the  the semileptonic $B_s \to \phi \ell^+\ell^-$
and present the theoretical predictions for many physical observables:
\begin{enumerate}
\item[(1)]
For  $B_s \to \phi  \ell^+\ell^-$  decays,  we  treat them as  a four body decay  $B_s \to \phi(\to K^- K^+)  \ell^+\ell^-$ described by  four kinematic
variables: the lepton invariant mass squared $q^2$ and three angles $(\thK, \thl,\Phi)$.
We defined and calculated the full angular decay distribution, the transverse amplitudes, the partially integrated decay amplitudes over the angles $(\thK,\thl,\Phi)$,
the CP averaged differential branching, the ratios $R^{e\mu}(\phi)$ and  $R^{\mu \tau}(\phi)$ of the branching ratios,
the forward-backward asymmetry $A_{FB}(q^2)$,  the $\phi$ polarization fraction $F_{L}(q^2)$,  the CP averaged (asymmetry) angular coefficients $S_{i}$ ($A_i$)
and the optimized observables $P_i$ and $P_i^\prime$.
Following Ref.~\cite{Doring2013} , where the authors approved that  the possible S-wave correction to the branching fractions of $B_s \to \phi (\to K^- K^+)  \ell^+\ell^-$ decays is small
and may modify the differential decay widths by about  $5 \%$ only,
we therefore will take the S-wave correction to the branching fractions as an additional uncertainty of $5\%$ in magnitude.

\item[(2)]
We used both the PQCD factorization approach and the ``PQCD+Lattice" approach to determine the values and their $q^2$-dependence of
the $B_s\to \phi$ transition form factors. We used the $z$-series parametrization to make the extrapolation for all form factors
from the low $q^2$ region to the endpoint $q^2_{max}$.
We will calculate the branching ratios and all other physical observables by using  the PQCD approach
itself  and the ``PQCD+Lattice'' approach respectively,  and compare their predictions with those currently available experimental measurements.

\end{enumerate}

The paper is organized as follows: In Sec.~\ref{sec:2}, we give a short review for the kinematics of the $B_s \to \phi \ell^+ \ell^-$
decays including distribution amplitudes of $B_s$ and $\phi$ mesons, and the  effective Hamiltonian for the quark level $b \to s \ell^+ \ell^-$.
In Sec.~\ref{sec:3}, we define explicitly all  physical observables for  $B_s \to \phi \ell^+ \ell^-$  decays.
In Sec.~\ref{sec:4} we present  our theoretical predictions  of all relevant physical observables of the considered decay modes, compare these predictions
with those currently available experimental measurements and make some phenomenological analysis. A short summary  is given in the last section.


\section{Kinematics and theoretical framework}\label{sec:2}

\subsection{ Kinematics and wave functions} \label{sec:2a}

We treat  the $B_s$ meson at rest as a heavy-light system. The kinematics of the semileptonic  $B_s \to \phi \ell^+ \ell^-$ decays
in the large-recoil (low $q^2$) region will be discussed below, where the PQCD factorization approach is applicable to the considered decays.
In the rest frame of $\bar{B}^0_s$ meson, we define the $\bar{B}^0_s$ meson momentum $p_1$, the $\phi $ momentum $p_2$ in the light-cone coordinates
as Ref.~\cite{Fan:2013qz}. We also use $x_i$ to denote the momentum fraction of light anti-quark in each meson and set the momentum
$p_i$ and $k_i$ ( the momenta carried by the spectator quark in $B_s$ and $\phi$ meson ) in the following forms:
\beq
\label{eq-mom-p1p2}
p_1&=&\frac{m_{B_s}}{\sqrt{2}}(1,1,0_\bot),\quad p_2=\frac{rm_{B_s}}{\sqrt{2}} (\eta^+,\eta^-,0_\bot),\non
k_1&=&(0,x_1\frac{m_{B_s}}{\sqrt{2}},k_{1\bot}), \quad k_2=\frac{m_{B_s}}{\sqrt{2}}(x_2r\eta^+,x_2r\eta^-,k_{2\bot}).
\eeq
 where the mass ratio $r=m_{\phi}/m_{B_s}$ , and the factor $\eta^\pm$ is defined in the following form:
\beq
\eta^\pm = \eta \pm \sqrt{\eta^2-1}, \quad {\rm with} \quad   \eta =\frac{1}{2r}\left [ 1+r^2-\frac{q^2}{m_{B_s}^2}\right],
\label{eq:eta9}
\eeq
where $q=p_1-p_2$ is the lepton-pair four-momentum.  For the final state $\phi$ meson, its longitudinal  and transverse polarization vector
$\epsilon_{L,T}$ can be written in the form of $ \epsilon_L=(\eta^+,-\eta^-,0_\bot)/\sqrt{2}$ and $\epsilon_T=(0,0,1)$.

For the $B_s$ meson wave function, we use the same kind of parameterizations as  in Refs.~\cite{Ali:2007ff,Xiao:2011tx,Wang:2012ab}
\beq
\Phi_{B_s}=\frac{i}{\sqrt{2N_{c}}} (\psl_{B_s} +m_{B_s}) \gamma_5 \phi_{B_s} ({k_1}).
\label{eq:bmeson}
\eeq
Here only the contribution of the Lorentz structure $\phi_{B_s} (k_1)$ is taken into account, since the contribution of the second Lorentz structure $\bar{\phi}_{B_s}$
is numerically small  \cite{Kurimoto:2001zj,Lu2003a} and has been neglected. We adopted the  distribution amplitude of the $B_s$  meson  in the similar form
as that of $B$-meson in the $SU(3)_f$ limit being widely used in the PQCD approach ~\cite{Ali:2007ff,Xiao:2011tx,Wang:2012ab}
\beq
 \phi_{B_s}(x,b)&=& N_{B_s} x^2(1-x)^2\mathrm{\exp} \left[ -\frac{m_{B_s}^2\; x^2}{2 \omega_{B_s}^2} -\frac{1}{2} (\omega_{B_s} b)^2\right].
\label{eq:phib}
\eeq
 In order to estimate  the theoretical uncertainties induced by the variations of $\phi_{B_s}(x,b)$, one usually
 take ~$\omega_{B_s}  =0.50 \pm 0.05$~GeV for $B_s^0$ meson \cite{li2003,Ali:2007ff}.
 The normalization factor ~$N_{B_s}$ depends on the values of the shape
 parameter $\omega_{B_s}$ and the decay constant $f_{B_s}$ and defined through the normalization relation : $\int_0^1dx\; \phi_{B_s}(x,0)=f_{B_s}/(2\sqrt{6})$ \cite{Ali:2007ff,Wang:2012ab}.

 For the vector meson $\phi$,  the longitudinal and transverse  polarization components can both provide the
 contribution. Here we adopt the wave functions of the vector $\phi $  as in Ref.~\cite{Ali:2007ff}:
\beq
\Phi_{\phi}^{||}(p,\epsilon_L)&=& \frac{i}{\sqrt{6}} \left [\not\! \epsilon_L m_{\phi}\phi_{\phi}(x)+\not\! \epsilon_L\psl\phi^{t}_{\phi}(x)
+m_{\phi}\phi^{s}_{\phi}(x)\right], \label{eq:phiv}\\
\Phi_{\phi}^{\perp}(p,\epsilon_T)&=& \frac{i}{\sqrt{6}} \left [\not\! \epsilon_T m_{\phi}\phi^{v}_{\phi}(x)+\not\! \epsilon_T\psl\phi^{T}_{\phi}(x)
+m_{\phi}i\epsilon_{\omega\upsilon\rho\sigma}\gamma_5\gamma^{\omega}\epsilon^{v}_{T}n^{\rho}v^{\sigma}\phi^{a}_{\phi}(x)\right],
\label{eq:phivT}
\eeq
where $p$ and $m_{\phi}$ are the momentum and the mass of the $\phi$ meson,  $ \epsilon_L $ and $ \epsilon_T $
correspond to the longitudinal and transverse polarization vectors of the vector meson $\phi$, respectively.
The  twist-2  DAs $\phi_{\phi}$ and $\phi_{\phi}^T$  in Eqs.~(\ref{eq:phiv},\ref{eq:phivT}) can be reconstructed as a Gegenbauer expansion   \cite{Ali:2007ff}:
\beq
\phi_{\phi}(x)&=&\frac{3f_{\phi}}{\sqrt{6}} x (1-x)\left[1+ \sum^2_{n=1}a_{n\phi}^{||}C_n^{3/2} (t)\right],\non
\phi^T_{\phi}(x)&=&\frac{3f^T_{\phi}}{\sqrt{6}} x (1-x)\left[1+ \sum^2_{n=1}a_{n\phi}^{\perp}C_n^{3/2} (t)\right], \label{eq:t2-02}
\eeq
where  $t=2x -1$, $a^{||,\perp}_{1,2}$ are the Gegenbauer moments, while $C^{3/2}_{1,2}$  are the Gegenbauer polynomials as given in Ref.~\cite{Ali:2007ff}.
$f_\phi$ and $f_\phi^T$ are the longitudinal and transverse components of the decay constants of the vector meson $\phi$ with $f_{\phi}=0.231\pm0.004$ GeV and $f^T_{\phi}=0.20\pm0.01$ GeV
as given in Ref.~\cite{Ali:2007ff}.  For  the relevant Gegenbauer moments we use the same ones as those in Refs.~\cite{Ball:2006wn,Ali:2007ff,xiao18a,xiao18b}.
\beq
a^{||,\perp}_1=0,\quad a_{2\phi}^{||}=0.18\pm0.08,\quad a_{2\phi}^{\perp}=0.14\pm0.07. \label{eq:gb02}
\eeq
The twist-3 DAs $\phi^{s,t}_{\phi}$ and $\phi^{v,a}_{\phi}$   in Eqs.~(\ref{eq:phiv},\ref{eq:phivT})  are the same ones as those defined in Ref.~\cite{Ali:2007ff}:
\beq
\phi^t_{\phi} = \frac{3f^T_{\phi}}{2\sqrt 6} t^2,   \quad \phi^s_{\phi}=\frac{3f_{\phi}^T}{2\sqrt 6} (-t)~,\quad
\phi^v_{\phi} =  \frac{3f_{\phi}}{8\sqrt 6}(1+t^2),   \quad \phi^a_{\phi}=\frac{3f_{\phi}}{4\sqrt 6} (-t)~, \label{eq:t3-01}
\eeq
where $t=2x-1$.

\subsection{Effective Hamiltonian for \texorpdfstring{$b \to s \ell^+ \ell^-$}{} decays }\label{sec:2b}

The effective Hamiltonian for the considered semileptonic decay $B_s\to \phi \ell^+ \ell^-$   is defined by  the same one as in Refs.~\cite{Buchalla:1995vs,Li:2008tk,Kindra:2018ayz,Singh:2019hvj,Nayek:2018rcq}:
\beq
{\cal H}_{\text{eff}} &=& -\frac{4G_F}{\sqrt{2}} \Big\{   V_{tb}V^*_{ts}  \left [ C_1(\mu)\mathcal{O}_1^c(\mu)+ C_2(\mu) \mathcal{O}_2^c(\mu)+ \sum_{i=3}^{10}{C}_i(\mu){\mathcal{O}}_i(\mu) \right] \non
&& + V_{ub}V^*_{us} \Big [     C_1(\mu)\left[\mathcal{O}_1^c(\mu)-\mathcal{O}^u_1(\mu)\right]  + C_2(\mu)\left[ \mathcal{O}_2^c(\mu)-\mathcal{O}^u_2(\mu) \right] \Big ] \Big \} + {\rm h.c}.,
\label{eq:heff}
\eeq
where  $G_F=1.16638\times 10^{-5}{\rm GeV}^{-2}$  is the Fermi constant,  $V_{ij}$ are the CKM matrix elements. For the operators ${\mathcal{O}}_i$ we adopt those as defined
in the so-called $\gamma_5$-free basis ~\cite{Chetyrkin:1996vx,Chetyrkin:1997gb}. Following Ref.~\cite{Gambino:2003zm}, the operators $\mathcal{O}_i$ can be written in the following form:
\begin{align}
{\mathcal{O}}^c_{1} &=(\bar s \gamma_{\mu} T^a P_L c)(\bar c \gamma_{\mu} T^a P_L b),&
{\mathcal{O}}^c_{2} &=(\bar s \gamma_{\mu} P_L c)(\bar c \gamma_{\mu} P_L  b),\non
{\mathcal{O}}^u_{1} &=(\bar s \gamma_{\mu} T^a P_L u)(\bar u \gamma_{\mu} T^a P_L b),&
{\mathcal{O}}^u_{2} &=(\bar s \gamma_{\mu} P_L u)(\bar u \gamma_{\mu} P_L b),\non
{\mathcal{O}}_{3} &=(\bar s \gamma_{\mu} P_L b){\textstyle \sum_q}(\bar q \gamma_{\mu}  q),&
{\mathcal{O}}_{4} &=(\bar s \gamma_{\mu} T^a P_L b) {\textstyle \sum_q}(\bar q \gamma_{\mu} T^a  q),\non
{\mathcal{O}}_{5} &=(\bar s \gamma_{\mu}\gamma_{\nu}\gamma_{\rho} P_L b){\textstyle \sum_q}(\bar q \gamma^{\mu}\gamma^{\nu}\gamma^{\rho}  q),&
{\mathcal{O}}_{6} &=(\bar s \gamma_{\mu}\gamma_{\nu}\gamma_{\rho} T^a P_L b){\textstyle \sum_q}(\bar q \gamma^{\mu}\gamma^{\nu}\gamma^{\rho} T^a q),\non
{\mathcal{O}}_{7} &=\frac{e}{g^2}m_b \bar (s \sigma^{\mu\nu} P_R b) F_{\mu\nu},&
{\mathcal{O}}_{8} &=\frac{1}{g}m_b \bar (s \sigma^{\mu\nu} T^a P_R b) G^a_{\mu\nu},\non
{\mathcal{O}}_{9} &= \frac{e^2}{g^2}(\bar s \gamma_{\mu} P_L b){\textstyle \sum_\ell}(\bar \ell  \gamma^{\mu} \ell), &
{\mathcal{O}}_{10}& =\frac{e^2}{g^2}(\bar s \gamma_{\mu} P_L b){\textstyle \sum_\ell}(\bar \ell  \gamma^{\mu} \gamma_5 \ell).\label{eq:operators4}
\end{align}
where ${\mathcal{O}}^{c,u}_{1,2}$ are the current-current operators,   ${\mathcal{O}}_{3-6}$ are the QCD penguin operators, ${\mathcal{O}}_{7,8}$ are the electromagnetic and chromomagnetic penguin operators respectively,
and finally  ${\mathcal{O}}_{9,10}$ are the semileptonic operators.  The inclusion of the factors $4\pi/g^2=1/{\alpha_s}$ in the definition of the operators ${\mathcal{O}}_{7,8,9,10}$ serves to
allow a more transparent organisation of the expansion of the relevant Wilson coefficients as  defined in Refs.~\cite{Bobeth:1999mk,Gambino:2003zm} up to next-to-next-to leading order (NNLO).
They are then evolved from the scale $\mu=m_W$ down to the scale $\mu=m_b$ using the renormalization group equations.


Since the contributions from the subleading chromomagnetic penguin, quark-loop and annihilation diagrams are highly suppressed for the considered  $b\to s \ell^+\ell^-$  decays \cite{Kindra:2018ayz}, we will neglect them in our calculations.
Using the effective Hamiltonian in Eq.~(\ref{eq:heff}), the decay amplitude for $b\to s \ell^+\ell^-$ loop transition can be decomposed as
a product of a short-distance contributions through Wilson coefficients and long-distance contribution which is further expressed in terms of form factors,
 \begin{eqnarray}
 {\cal A}(b\to s\ell^+ \ell^-)=\frac{G_F}{\sqrt{2}}\frac{\alpha_{\rm{em}}}{\pi}V_{tb}V^*_{ts}\bigg\{
 &&C_9^{\rm{eff}}(q^2)
 [\bar s \gamma_{\mu} P_L b][\bar \ell\gamma^{\mu}\ell] + C_{10}[\bar s\gamma_{\mu} P_L b]
 [\bar \ell\gamma^{\mu}\gamma_5\ell]\non
 &&- 2m_bC_7^{\rm{eff}}\big[\bar s i\sigma_{\mu\nu}\frac{q^{\nu}}{q^2}  P_R b\big][\bar \ell\gamma^{\mu}\ell] \bigg\},\label{eq:Ampbtos}
 \end{eqnarray}
where $C_7^{eff}(\mu)$ and $C_9^{eff}(\mu)$ are the effective Wilson coefficients, defined as in Refs.~\cite{Chen:2001zc,Wang:2012ab}
\beq
C_7^{\rm{eff}}(\mu)&=&C_7(\mu)+C^{\prime}_{b\to s\gamma}(\mu), \label{eq:c7eff}\\
C_9^{\rm{eff}}(\mu,q^2)&=&C_9(\mu)+Y_{\rm{pert}}(q^2)+Y_{\rm{res}}(q^2). \label{eq:c9eff}
\eeq
The term $C^{\prime}_{b \to s\gamma}$ in Eq.~(\ref{eq:c7eff})  is the absorptive part of $b\to s\gamma$ transition and was given in Ref.~\cite{Chen:2001zc}
\beq
C^{\prime}_{b \to s\gamma}(\mu)=i\alpha_s\left \{\frac{2}{9}\eta^{14/23}\left [  \frac{x_t \left( x_t^2-5x_t-2\right ) }{ 8 \left (x_t-1 \right )^3 }+\frac{3x_t^2 \ln x_t}{4(x_t-1)^4}  -0.1687 \right ]-0.03C_2(\mu) \right \},
\eeq
where  $x_t=m_t^2/m_W^2$ and $\eta=\alpha_s(m_W)/\alpha_s(\mu)$.
The explicit expressions of the term $Y_{\rm{pert}}(q^2)$  and  $Y_{\rm{res}}(q^2)$  in Eq.~(\ref{eq:c9eff}) are of the following form \cite{Ali:1991is,Lim:1988yu,Deshpande:1988bd,ODonnell:1991cdx,Nayek:2018rcq}
 \beq
Y_{\rm{pert}}(q^2) &=& 0.124\, \omega(\hat{s})+g(\hat{m}_c,\hat{s})C_0+\lambda_u\left [ g(\hat{m}_c,\hat{s})-g(\hat{m}_u,\hat{s}) \right] (3C_1+C_2)  \non
&&- \frac{1}{2}g(\hat{m}_b,\hat{s})(C_3+3C_4) - \frac{1}{2}g(\hat{m}_b,\hat{s})(4C_3+4C_4+3C_5+C_6) \non
&& + \frac{2}{9}(3C_3+C_4 +3C_5+C_6) , \label{eq:ypert}\\
Y_{\rm{res}}(q^2)&=&-\frac{3\pi}{\alpha_{\rm{em}}^2} \Big[  C_0 \cdot\! \sum_{V=J/\Psi,\Psi'...}\frac{m_V {\cal B}(V\to l^+l^-)\Gamma_{\rm{tot}}^V} {q^2-m_V^2+im_V\Gamma_{\rm{tot}}^V}\non
&&- \lambda_u \; g(\hat{m}_u,\hat{s})(3C_1+C_2)\cdot \sum_{V=\rho,\omega,\phi}\frac{m_V {\cal B}(V\to l^+l^-)
\Gamma_{\rm{tot}}^V}{q^2-m_V^2+im_V\Gamma_{\rm{tot}}^V}\Big],\label{eq:3-31}
\eeq
where $C_0=3C_1+C_2+3C_3+C_4+3C_5+C_6$, $\hat{s}\!=\!q^2/m^2_b$, $\hat{m}_q\!=\!{m_q}/{m_b}$  and the CKM ratio $\lambda_u=V_{ub}V^*_{us}/(V_{tb} V^*_{ts})$.
In Eq.~(\ref{eq:ypert}), the function $\omega(\hat{s})$ is the soft-gluon correction to the matrix element of operator $\mathcal{O}_{9}$.
The function $g(\hat{m}_q,\hat{s})$ in Eqs.~(\ref{eq:ypert},\ref{eq:3-31})  is related to the basic fermion loop.
The contributions from four-quark operators ${\mathcal{O}}_{1}-{\mathcal{O}}_{6}$ are usually combined with coefficient $C_9$ into an "effective" one.
One can find the explicit expressions of the function $\omega(\hat{s})$ and $g(\hat{m}_q,\hat{s})$  easily for example in Ref.~\cite{Jin:2020jtu} and references therein.

The term  $Y_{\rm{pert}}(q^2)$ in Eqs.~(\ref{eq:c9eff},\ref{eq:ypert}) defines the short distance perturbative part that involves the indirect contributions from the matrix element of the four quark operators $\sum^{10}_{i=1}\langle\ell^+\ell^-s|\mathcal{O}_i|b\rangle$  \cite{Ali:1991is,Lim:1988yu,Deshpande:1988bd,ODonnell:1991cdx}  and lies at the place far away from $c\bar c$ resonance regions.

The term $Y_{\rm{res}}(q^2)$ in  Eqs.~(\ref{eq:c9eff},\ref{eq:3-31}) describes  the long distance  resonant  contributions related with the $B_s \to \phi  V \to \phi  (V\to l^+l^-) $  transitions
in the resonance regions,  where $V=(\rho,\omega,\phi,\jpsi, \psi',\cdots )$  are  the light vector mesons and $c\bar{c}$ charmonium states.
Up to now the term $Y_{\rm{res}}(q^2)$ can not be calculated from the first principle of QCD and may also introduce the double-counting problem with the term $Y_{\rm{pert}}(q^2)$.
For more details about such kinds of  double-counting problem, one can see the discussions as given in Refs.~\cite{Khod2010,Khod2013}.
In this paper, we checked the possible effects on the theoretical predictions for the branching ratios and other considered physical observables by including the term
$Y_{\rm{res}}(q^2)$ or not in our numerical calculations, and we found that the resulted variations of the theoretical predictions are less than $5\%$.
It is much smaller than the total theoretical errors: say around  $30-40\%$.  According to the argument in Ref.~\cite{Ahmed:2013lba},  the term $Y_{\rm{res}}(q^2)$  is also generally small.
Because of its smallness and the possible  double-counting problem we here simply drop the term  $Y_{\rm{res}}(q^2)$  out in our numerical evaluations for all physical observables considered
in this paper.

\subsection{    \texorpdfstring{$B_s \to\phi$ }{} transition form factors }\label{sec:2c}

For the vector meson $\phi$ with polarization vector $\epsilon^{*}$,  as usual, the relevant form factors for $B_s\to \phi $  transitions are
$V(q^2)$  and $A_{0,1,2}(q^2)$ of the vector and axial-vector currents, and $T_{1,2,3}$ of the tensor currents.
Between the form factors  $A_{0,1,2}(q^2)$ at the point $q^2=0$, there is an exact relation $2m_{\phi} A_0(0) =(m_{B_s}+ m_{\phi})A_1(0)-(m_{B_s}- m_{\phi})A_2(0) $
in order to avoid  the kinematical singularity.
Between  the form factor $T_{1,2}$, there also exist a relation  $T_1(0)=T_2(0)$  in an algebraic manner which is implied by the identity
$\sigma^{\mu\nu}\gamma_5=-\frac{i}{2}\epsilon^{\mu\nu\alpha\beta} \sigma_{\alpha\beta}$ with the $\epsilon^{0123}=+1$ convention for the the Levi-Civita tensor.

Using the well-studied wave functions as given in previous subsection, the PQCD factorization formulas for the relevant form factors of $B_s \to \phi \ell^+\ell^-$ decays
can be calculated and  written in the following form:
\beq
V(q^2)&=& 8\pi m_{B_s}^2C_F(1\!+\!r)\int dx_1 dx_2\int b_1 db_1 b_2 db_2\phi_{B_s}(x_1)\non
&& \times \Bigl\{\Big[\!-\!x_2r\phi^v_{\phi}(x_2)\!+\!\phi^T_{\phi}(x_2)\!+\!\frac{1\!+\!x_2r\eta}{\sqrt{\eta^2\!-\!1}}\phi^a_{\phi}(x_2)\Big] \!\cdot \!H_1(t_1)\non
&& \!+\!\Big[ \left (r\!+\!\frac{x_1}{2\sqrt{\eta^2\!-\!1}} \right ) \phi^v_{\phi}(x_2)
\!-\!\frac{x_1-2r\eta}{2\sqrt{\eta^2\!-\!1}}\phi^a_{\phi}(x_2)\Big]\!\cdot \!H_2(t_2)\Bigr \}, \label{eq:Vq2}
\eeq
\beq
A_0(q^2)&=& 8\pi m_{B_s}^2C_F\int dx_1 dx_2\int b_1 db_1 b_2 db_2\phi_{B_s}(x_1) \times \Bigl\{\Big[ \left (1\!+\!x_2r(2\eta\!-\!r) \right )\phi_{\phi}(x_2)\non
&& +\!(1\!-\!2x_2)r\phi^t_{\phi}(x_2)\!
+\!\frac{(1\!-\!r\eta)-2x_2r (\eta\!-\!r)}{\sqrt{\eta^2\!-\!1}}\phi^s_{\phi}(x_2)\Big]\!\cdot \!H_1(t_1)\non
&& \!+\!\Big[\Big[\frac{x_1}{\sqrt{\eta^2\!-\!1}} \left (\frac{\eta\!+\!r}{2}\!-\!r\eta^2 \right )\!+
\!\left (\frac{x_1}{2}\!-\!x_1r\eta\!+\!r^2 \right ) \Big]\phi_{\phi}(x_2)\non
&& -\!\Big [ \frac{x_1(1\!-\!r\eta)+2r(r\!-\!\eta)}{\sqrt{\eta^2\!-\!1}}\!-\!x_1r\Big]\phi^s_{\phi}(x_2)\Big ]
\!\cdot \!H_2(t_2)\Bigr \},\label{eq:A0q2}
\eeq
\beq
A_1(q^2)&=&16\pi m_{B_s}^2C_F\frac{r}{1\!+\!r}\int dx_1 dx_2\int b_1 db_1 b_2 db_2\phi_{B_s}(x_1)\non
&& \times \Bigl\{\Big[ (1\!+\!x_2r\eta)\phi^v_{\phi}(x_2)
\!+\! (\eta\!-\!2x_2r)\phi^T_{\phi}(x_2)\!+\!x_2r\sqrt{\eta^2\!-\!1}\phi^a_{\phi}(x_2)\Big]\!\cdot \!H_1(t_1)\non
&& \!+\!\Big[\left (r\eta\!-\!\frac{x_1}{2} \right )\phi^v_{\phi}(x_2)\!+\!\left (r\sqrt{\eta^2\!-\!1}\!+\!\frac{x_1}{2}\right )
\phi^a_{\phi}(x_2)\Big]\!\cdot \!H_2(t_2)\Bigr \},\label{eq:A1q2}
\eeq
\beq
A_2(q^2)&=&\frac{(1\!+\!r)^2(\eta\!-\!r)}{2r(\eta^2\!-\!1)}A_1(q^2)-8\pi m_{B_s}^2C_F\frac{1\!+\!r}{\eta^2\!-\!r}\int dx_1 dx_2\int b_1 db_1 b_2 db_2\phi_{B_s}(x_1)\non
&& \times \Bigl\{\Big[\big[\eta \left (1\!-\!x_2r^2\right )\!+\!r \left (x_2(2\eta^2\!-\!1)\!-\!1 \right )\big]\phi_{\phi}(x_2)
\!+\!\Big[1\!+\!2x_2r^2\!-\!(1\!+\!2x_2)r\eta\Big] \phi^t_{\phi}(x_2)\non
&& +r(1\!-\!2x_2)\sqrt{\eta^2\!-\!1}\phi^s_{\phi}(x_2)\Big]\!\cdot \!H_1(t_1)\non
&& +\!\Big[\Big[ \left ( r\eta\!-\!\frac{1}{2} \right )x_1\sqrt{\eta^2\!-\!1}\!-\!\Big [ r \left (r\eta\!-\!1\!-\!x_1\eta^2\right )\!+\!\frac{x_1(r\!+\!\eta)}{2}\Big] \Big] \phi_{\phi}(x_2)\non
&&+\!\Big[  x_1(r\eta\!-\!1)\!+\!(x_2\!-\!2)r\sqrt{\eta^2\!-\!1}\Big] \phi^s_{\phi}(x_2)\Big]\!\cdot \!H_2(t_2)\Bigr \}, \label{eq:A2q2}
\eeq
\beq
T_1(q^2)&=&8\pi m_{B_s}^2C_F\int dx_1 dx_2\int b_1 db_1 b_2 db_2\phi_{B_s}(x_1)\times \Bigl\{\Big[(1\!-\!2x_2)r\phi^v_{\phi}(x_2)\! \non
&&+ \left (1\!+\!2x_2r\eta\!-\!x_2r^2 \right )
\phi^T_{\phi}(x_2)\!+\!\frac{1\!+\!2x_2r^2\!-\!(1\!+\!2x_2)r\eta}{\sqrt{\eta^2\!-\!1}}\phi^a_{\phi}(x_2)\Big ]\!\cdot \!H_1(t_1)\non
&&+  \Big [ \Big[ \left (1\!-\!\frac{x_1}{2} \right )r-\frac{x_1(r\eta\!-\!1)}{2\sqrt{\eta^2\!-\!1}}\Big ]\phi^v_{\phi}(x_2)\non
&& +\Big[ \frac{r(\eta\!-\!r)}{\sqrt{\eta^2\!-\!1}}\!+\!\frac{x_1}{2}\left (r+\frac{r\eta\!-\!1}{\sqrt{\eta^2\!-\!1}}\right ) \Big ] \phi^a_{\phi}(x_2)\Big ] \cdot H_2(t_2)\Bigr \},\quad
\label{eq:T1q2}
\eeq
\beq
T_2(q^2)&=&16\pi m_{B_s}^2C_F\frac{r}{1\!-\!r^2}\int dx_1 dx_2\int b_1 db_1 b_2 db_2\phi_{B_s}(x_1)\non
&& \times \Bigl\{\Big[(1\!-\!(1+2x_2)r\eta\!+\!2x_2r^2)\phi^v_{\phi}(x_2)\!\non
&& +\! \Big (x_2r\eta(2\eta\!-\!r)\!-\!x_2r\!+\!\eta\!-\!r\Big )\phi^T_{\phi}(x_2)
+\!(1\!-\!2x_2)r\sqrt{\eta^2\!-\!1}\phi^a_{\phi}(x_2)\Big]\!\cdot \!H_1(t_1)\!\non
&& +\!\Big[\Big[\frac{x_2}{2}\Big (1\!+\!\frac{\eta}{\sqrt{\eta^2\!-\!1}} \Big) (r\eta\!-\!1)\!+
\!\Big ( r\!+\!\frac{x_1}{2\sqrt{\eta^2\!-\!1}} \Big) (\eta\!-\!r)\Big]\phi^v_{\phi}(x_2)\non
&& \!+\!\Big[ \Big (1\!-\!\frac{x_1}{2} \Big)r\sqrt{\eta^2-1}\!+\!\frac{x_1}{2}(1\!-\!r\eta)\Big]\phi^a_{\phi}(x_2)\Big]
\!\cdot \!H_2(t_2)\Bigr \},\label{eq:T2q2}
\eeq
\beq
T_3(q^2)&=&\frac{(1\!-\!r)^2(\eta\!+\!r)}{2r(\eta^2\!-\!1)}T_2(q^2)-8\pi m_{B_s}^2C_F\frac{1\!-\!r^2}{\eta^2\!-\!1}\int dx_1 dx_2\int b_1 db_1 b_2 db_2\phi_{B_s}(x_1)\non
&&\hspace{-1cm}\times \Bigl\{\Big[\frac{\eta^2\!-\!(1\!+\!2x_2)r\eta\!+\!2x_2r^2}{\eta\!-\!r}\phi_{\phi}(x_2)
\!+\!(1\!+\!x_2r\eta)\phi^t_{\phi}(x_2)\!+\!x_2r\sqrt{\eta^2\!-\!1}\phi^s_{\phi}(x_2)\Big]\!\cdot \!H_1(t_1)\non
&&\hspace{-1cm} +\!\Big[\big[r\!-\!\frac{x_1}{2}(\eta\!+\!\sqrt{\eta^2\!-\!1})\big]\phi_{\phi}(x_2)
\!+\!(x_1\!+\!2r\sqrt{\eta^2\!-\!1})\phi^s_{\phi}(x_2)\Big]\!\cdot \!H_2(t_2)\Bigr \},\label{eq:T3q2}
\eeq
where $r=m_{\phi}/m_{B_s}$,  the twist-2 DAs $(\phi_{\phi},\phi^T_{\phi})$ and the  twist-3 DAs   $(\phi_{\phi}^{s,t},\phi^{v,a}_{\phi})$
have  been defined in Eqs.~(\ref{eq:t2-02},\ref{eq:t3-01}). The function $H_i(t_i)$ in above equations are of the following form
\beq
H_i(t_i)= h_i(x_1,x_2,b_1,b_2) \cdot \alpha_s(t_i) \cdot S_t(x_2) \exp \left[-S_{ab}(t_i)\right],\quad for \quad i=(1,2).
\label{eq:hiti}
\eeq
The hard functions $h_{1,2}(x_1,x_2,b_1,b_2)$  come form the Fourier transform of virtual quark and gluon propagators and they  can be defined by
\beq
h_1&=&K_0(\beta_1 b_1) \left [\theta(b_1-b_2)I_0(\alpha_1 b_2)K_0(\alpha_1 b_1) +\theta(b_2-b_1)I_0(\alpha_1 b_1)K_0(\alpha_1 b_2) \right ], \non
h_2&=&K_0(\beta_2 b_1) \left [\theta(b_1-b_2)I_0(\alpha_2 b_2)K_0(\alpha_2 b_1) +\theta(b_2-b_1)I_0(\alpha_2 b_1)K_0(\alpha_2 b_2) \right ],
\eeq
where $K_0$ and $I_0$ are modified Bessel functions, and
\beq
\alpha_1 = m_{B_s}\sqrt{x_2r \eta^+},\quad  \alpha_2=m_{B_s}\sqrt{x_1 r \eta^+ - r^2 +r_s^2},\quad
\beta_1 = \beta_2=m_{B_s}\sqrt{x_1x_2 r \eta^+},\quad
\eeq
where $r=m_{\phi}/m_{B_s}, r_s=m_s/m_{B_s}$.
The hard scales $t_i$  in Eq.~(\ref{eq:hiti})   are chosen as the largest scale of the virtuality of the internal particles in the hard $b$-quark decay diagram, including $1/b_i(i=1,2)$:
\beq
t_1=\max\{\alpha_1, 1/b_1, 1/b_2\},\quad t_2=\max\{\alpha_2,1/b_1, 1/b_2\}.
\eeq
The threshold resummation factor $S_t(x)$ in Eq.~(\ref{eq:hiti})  is adopted from \cite{Kurimoto:2001zj},
\beq
\label{eq-def-stx} S_t=\frac{2^{1+2c}\Gamma(3/2+c)}{\sqrt{\pi}\Gamma(1+c)}[x(1-x)]^c,
\eeq
with a fitted parameter $c(Q^2)=0.04 Q^2 -0.51 Q+1.87$~\cite{Li:2009pr} and $Q^2=m^2_{B_s}(1-r^2)$~\cite{Wang:2015uea}. The function $S_t(x)$ is  normalized to unity.
The function $\exp[-S_{ab}(t)]$  in Eq.~(\ref{eq:hiti})  contains the Sudakov logarithmic
corrections and the renormalization group evolution effects of both the wave functions and the hard scattering amplitude, for more details of function $\exp[-S_{ab}(t)]$
one can see Refs.~\cite{Lu:2000em,Kurimoto:2001zj}.

\section{Observables for \texorpdfstring{$B_s \to \phi \ell^+ \ell^-$}{mode} decays}{}\label{sec:3}

In experimental analysis, the $\bar{B}_s\to \phi \ell^+\ell^-$  decay is treated  as  the four body differential decay distribution $\bar{B}_s\to \phi (\to K^+ K^-)\ell^+\ell^-$,
and has  been described in terms of  the four kinematic variables \cite{Aaij:2015oid,DescotesGenon:2012zf,Descotes-Genon:2013vna,Bobeth:2008ij}:
the lepton invariant mass squared $q^2$ and the three decay angles $\vec{\Omega}=(\cos\thK, \cos\thl, \Phi)$.
The angle $\thK$ is the angle between the direction of flight of $K^+$ and $B_s$ meson in the rest frame of $\phi$,
$\thl$ is the angle made by $\ell^-$ with respect to the $B_s$ meson in the dilepton rest frame and $\Phi$ is the azimuthal angle between the two planes
formed by dilepton and $K^+ K^-$.

With the hadronic and leptonic amplitudes defined in Eq.~(\ref{eq:Ampbtos}), we write down the four fold differential distribution of four-body
$\bar{B}_s\to \phi(\to K^+ K^-)\ell^+\ell^-$ decay \cite{Kindra:2018ayz,Aaij:2015oid,Altmannshofer:2008dz,Becirevic:2011bp},
\begin{equation}
  \frac{d^4\Gamma}{dq^2\, d\vec{\Omega}} =
   \frac{9}{32\pi} I(q^2, \vec{\Omega}), \quad d\vec{\Omega}=d\cos\thK \, d\cos\thl\, d\Phi, \label{eq:d4Gamma}
\end{equation}
where  the functions  $ I(q^2, \vec{\Omega})$ can be written in terms of a set of angular coefficients and trigonometric functions ~\cite{Altmannshofer:2008dz}:
\beq
 I(q^2, \vec{\Omega})& =& {\textstyle \sum_i} I_{i}(q^2){f_{i}(\vec{\Omega})} \non
    & =&       I_{1s} \sin^2\thK + I_{1c} \cos^2\thK       + (I_{2s} \sin^2\thK + I_{2c} \cos^2\thK) \cos 2\thl \non
    & +&  I_3 \sin^2\thK \sin^2\thl \cos 2\Phi       + I_4 \sin 2\thK \sin 2\thl \cos\Phi \non
    & + & I_5 \sin 2\thK \sin\thl \cos\Phi       + I_{6s} \sin^2\thK \cos\thl       + I_7 \sin 2\thK \sin\thl \sin\Phi  \non
    & + & I_8 \sin 2\thK \sin 2\thl \sin\Phi       + I_9 \sin^2\thK \sin^2\thl \sin 2\Phi\,.  \label{eq:angulardist}
\eeq
For the CP-conjugated mode $B_s\to \phi(\to K^- K^+)\ell^+\ell^-$, the corresponding expression of the angular decay distribution is
\beq
  \frac{d^4\bar{\Gamma}}{dq^2\, d\vec{\Omega}} =    \frac{9}{32\pi} \bar{I}(q^2, \vec{\Omega})\,, \label{eq:d4Gammabar}
\eeq
where  the function $ \bar{I}(q^2, \vec{\Omega})$ is obtained from    $I(q^2, \vec{\Omega})$  in Eq.~(\ref{eq:angulardist}) by making the complex conjugation
for all weak phases in $I_i$ ~\cite{Altmannshofer:2008dz}, and numerically by the following substitution:
\beq
 I_{1(c,s),2(c,s),3,4,7} \to \bar{I}_{1(c,s),2(c,s),3,4,7}, \quad I_{5,6s,8,9} \to -\bar{I}_{5,6s,8,9}.\label{eq:replacement}
\eeq
The minus sign in Eq.~(\ref{eq:replacement}) is a result of the convention that, under the previous definitions of three angles $(\theta_K, \theta_l, \Phi)$,
a CP transformation interchanges  the lepton and anti-lepton, i.e., leading to the transformation $\thl \to \thl-\pi$ and $\Phi \to -\Phi$.

The angular coefficients $ I_i$, which are functions of $q^2$ only, are usually expressed  in terms of the transverse amplitudes~\cite{Matias:2012xw,Aaij:2015oid}.
In the limit of massless leptons, there are six such complex amplitudes: $\cala_0^{L,R}$, $\cala_\|^{L,R}$ and $\cala_\perp^{L,R}$, where $L$ and $R$ refer to the chirality of
the leptonic current. For the massive case, an additional complex amplitude $\cala_t$ is required, where the timelike component of the virtual gauge boson (which
can later decay into dilepton) couple to an axial-vector current.

In Table \ref{tab:table1},  we summarize the treatment of the angular distribution by decomposition of the angular coefficients $I_i(q^2)$ into seven transverse amplitude
$\cala^{L,R}_{\perp,\|,0}$ and $\cala_t$  as well as the corresponding trigonometric factor ${f_{i}(\vec{\Omega})}$.
Here we will not consider scalar contribution to facilitate the comparison with Ref.~\cite{Jin:2020jtu}.
Notice that the distribution including lepton masses (but neglecting scalar $I_{6c}=0$) contains eleven  $I_i$  where only 10 of them are independent \cite{Egede:2010zc,Matias:2012xw}.
In the limit of massless leptons, it is easy to obtain the relations $I_{1s}=3I_{2s}$ and $I_{1c}=-I_{2c}$ ~\cite{Altmannshofer:2008dz}.

\begin{table}[thb]
\caption{The explicit expressions of the angular coefficients $I_i(q^2)$ and $  f_{i}(\vec{\Omega})$ appeared in Eq.~(\ref{eq:angulardist}).  }
\label{tab:table1}
\begin{center}
\begin{tabular}{|l|l|l|}
\hline
$i$ & $I_{i}(q^2)$ & $ f_{i}(\vec{\Omega})$ \\
\hline
$1s$ & $(\frac{3}{4}\!-\!\hat{m}^2_\ell)\left[ |{\cal A}_{\parallel}^{\rm L}|^{2}\!+\! |{\cal A}_{\perp}^{\rm L}|^{2} \!+\!|{\cal A}_{\parallel}^{\rm R}|^{2} \!+\! |{\cal A}_{\perp}^{\rm R}|^{2}\right]\!+\!4 \hat{m}^2_\ell {\rm Re}\left[ {\cal A}_{\perp}^{\rm L}{\cal A}_{\perp}^{\rm R\ast}\! +\!{\cal A}_{\parallel}^{\rm L}{\cal A}_{\parallel}^{\rm R\ast}\right] $ & $\sin^{2}\thK$ \\
$1c$ &  $|{\cal A}_{0}^{\rm L}|^{2} + |{\cal A}_{0}^{\rm R}|^{2} +4 \hat{m}^2_\ell \left[|{\cal A}_{t}|^{2}+2{\rm Re}[{\cal A}_{0}^{\rm L}{\cal A}_{0}^{\rm R\ast}]\right]$ & $\cos^{2}\thK$ \\
$2s$ & $\frac{1}{4} \beta^2_\ell \left[ |{\cal A}_{\parallel}^{\rm L}|^{2} + |{\cal A}_{\perp}^{\rm L}|^{2}  + |{\cal A}_{\parallel}^{\rm R}|^{2} + |{\cal A}_{\perp}^{\rm R}|^{2}\right]$ & $\sin^{2}\thK\cos 2\thl$ \\
$2c$ &  $-\beta^2_\ell \left[|{\cal A}_{0}^{\rm L}|^{2} +  |{\cal A}_{0}^{\rm R}|^{2}\right]$ & $\cos^{2}\thK\cos 2\thl$ \\
3 & $\frac{1}{2} \beta^2_\ell \left[ |{\cal A}_{\perp}^{\rm L}|^{2} - |{\cal A}_{\parallel}^{\rm L}|^{2}  + |{\cal A}_{\perp}^{\rm R}|^{2} - |{\cal A}_{\parallel}^{\rm R}|^{2} \right]$  & $\sin^{2}\thK \sin^{2}\theta_{\ell} \cos 2\Phi$ \\
4 & $ \sqrt{\frac{1}{2}} \beta^2_\ell {\rm Re}({\cal A}_{0}^{\rm L} {\cal A}_{\parallel}^{{\rm L}\ast} + {\cal A}_{0}^{\rm R} {\cal A}_{\parallel}^{{\rm R}\ast}) $  & $\sin 2\thK \sin 2\thl \cos \Phi$ \\
5 &  $\sqrt{2} \beta_\ell {\rm Re}({\cal A}_{0}^{\rm L} {\cal A}_{\perp}^{{\rm L}\ast} - {\cal A}_{0}^{\rm R} {\cal A}_{\perp}^{{\rm R}\ast})$ &  $\sin 2\thK \sin \thl \cos \Phi$ \\
$6s$ &  $2 \beta_\ell {\rm Re}({\cal A}_{\parallel}^{\rm L} {\cal A}_{\perp}^{{\rm L}\ast} - {\cal A}_{\parallel}^{\rm R} {\cal A}_{\perp}^{{\rm R}\ast})$ &  $\sin^{2}\thK \cos \thl$ \\
7 &  $\sqrt{2} \beta_\ell {\rm Im}({\cal A}_{0}^{\rm L} {\cal A}_{\parallel}^{{\rm L}\ast} - {\cal A}_{0}^{\rm R} {\cal A}_{\parallel}^{{\rm R}\ast})$ &  $\sin 2\thK \sin \thl \sin \Phi$ \\
8 &  $\sqrt{\frac{1}{2}} \beta^2_\ell {\rm Im}({\cal A}_{0}^{\rm L} {\cal A}_{\perp}^{{\rm L}\ast} + {\cal A}_{0}^{\rm R} {\cal A}_{\perp}^{{\rm R}\ast})$&  $\sin 2\thK \sin 2\thl \sin \Phi$ \\
9 &  $\beta^2_\ell {\rm Im}({\cal A}_{\parallel}^{{\rm L}\ast}{\cal A}_{\perp}^{\rm L}  +  {\cal A}_{\parallel}^{{\rm R}\ast}{\cal A}_{\perp}^{\rm R}) $ &  $\sin^{2}\thK \sin^{2}\thl \sin 2\Phi$ \\
\hline
\end{tabular}
\end{center}
\end{table}

The seven transverse  amplitudes  $\cala_0^{L,R}$, $\cala_\|^{L,R}$, $\cala_\perp^{L,R}$ and $\cala_t$ of $B_s \to \phi \ell^+\ell^-$ decay, in turn,  can be parameterized  by means of the relevant
form factors~\cite{Egede:2008uy,Altmannshofer:2008dz}:
\beq
{\cal A}_{\perp}^{\rm L,R} &=&- N_\ell\sqrt{2 N_{\phi}} \sqrt{\lambda} \left[(C_9^{\rm{eff}}\mp C_{10})\frac{V(q^2)}{m_{B_s}+m_{\phi}}+2 \hat{m}_b C_7^{\rm{eff}} T_1(q^2)\right], \label{eq:trans12}\\
{\cal A}_{\parallel}^{\rm L,R} &=&N_\ell\sqrt{2 N_{\phi}} \Big [ (C_9^{\rm{eff}}\mp C_{10})(m_{B_s}+m_{\phi})A_1(q^2)+2 \hat{m}_b C_7^{\rm{eff}} (m^2_{B_s}-m^2_{\phi})T_2(q^2) \Big ], \label{eq:trans34}
\eeq
\beq
{\cal A}_{0}^{\rm L,R} &=&\frac{N_\ell\sqrt{N_{\phi}}}{2m_{\phi} \sqrt{q^2}}\Big\{(C_9^{\rm{eff}}\!\mp\! C_{10})
\left [ (m^2_{B_s}\!-\!m^2_{\phi}\!-\!q^2)(m_{B_s}\!+\!m_{\phi})A_1(q^2) \!-\!\frac{\lambda}{m_{B_s}\!+\!m_{\phi}}A_2(q^2) \right ]\non
&&+2 m_b C_7^{\rm{eff}} \Big [ (m^2_{B_s}\!+\!3m^2_{\phi}\!-\!q^2)T_2(q^2)\!-\!\frac{\lambda}{m^2_{B_s}\!-\!m^2_{\phi}}T_3(q^2) \Big ]\Big\},  \label{eq:trans56}\\
{\cal A}_{t }&=&2 N_\ell\sqrt{N_{\phi}}\frac{\sqrt{\lambda}}{\sqrt{q^2}}C_{10}A_0(q^2),  \label{eq:trans7}
\eeq
where $\lambda = (m^2_{B_s}\!-\!m^2_{\phi}\!-\!q^2)^2\!-\!4m^2_{\phi} q^2$,   $\hat{m}_b=m_b/q^2$  and the normalization constants are given as:
 \beq
 N_\ell=\frac{i \alpha_{em} G_F}{4\sqrt{2}\pi}V_{tb}V^\ast_{ts},\quad
 N_{\phi}=\frac{8 \sqrt{\lambda}q^2}{3\times256\pi^3m^3_{B_s}}    \sqrt{1\!-\frac{4m_\ell^2}{q^2}} {\cal B}(\phi\to K^+K^-).
 \eeq
In numerical calculations, we take ${\cal B}(\phi\to K^+K^-)=0.492$ from PDG 2018 \cite{pdg2018}. It is easy to see that the narrow width approximation works well in the
case of $\phi$ meson  since $\Gamma_\phi/{m_\phi}=4.17\times10^{-3}\sim0$.

Analogous to Ref.~\cite{Altmannshofer:2008dz}, to separate CP-conserving and CP-violating effects,
one can define the CP averaged angular coefficients $S_i$ and CP asymmetry angular coefficients $A_i$ normalized by the differential (CP-averaged) decay rate to reduce the theoretical uncertainties,
\beq
S_i = \frac{I_i+\bar{I}_i}{d(\Gamma+\bar{\Gamma})/dq^2},\quad
A_i = \frac{I_i-\bar{I}_i}{d(\Gamma+\bar{\Gamma})/dq^2},
\label{eq:A&S}
\eeq
where $I_i$ and $\bar{I}_i$ have been defined in Eqs.~(\ref{eq:angulardist},\ref{eq:d4Gammabar},\ref{eq:replacement}) and Table \ref{tab:table1}, and
the differential decay rate reads (analogously for $\bar{\Gamma}$),
\beq
{\frac{d\Gamma}{dq^2}} = {\frac{1}{4}} \left( 3I_{1c} + 6I_{1s} - I_{2c} - 2I_{2s} \right) \,. \label{eq:dGdq2}
\eeq
Based on the definition of $S_i$ , one can find the relation $3 S_{1c}+6 S_{1s}-S_{2c}-2 S_{2s}=4$.
Consequently, all established observables can be expressed in terms of $S_i$ and $A_i$:
\begin{itemize}
\item[(1)] the CP asymmetry
\beq
A_{\rm CP}(q^2) = \frac{d\Gamma /dq^2 - d\bar{\Gamma}/dq^2}{ d\Gamma /dq^2 + d\bar{\Gamma}/dq^2 }= {\frac{1}{4}} \left( 3A_{1c} + 6A_{1s} - A_{2c} - 2A_{2s} \right).
\eeq
\item[(2)]
The lepton forward-backward (CP) asymmetry:
\beq
{A}_{\mathrm{FB}}(q^{2})=\frac{\left[\int_{0}^{1}-\int_{-1}^{0}\right] d \cos \theta_{\ell}\frac{d^{2}
(\Gamma-\bar{\Gamma})}{d q^{2} d \cos \theta_{\ell}} }{d( \Gamma+\bar{\Gamma}) / d q^{2}}=\frac{3}{4}S_{6s}, \\
{A}^{\rm CP}_{\mathrm{FB}}(q^{2})=\frac{\left[\int_{0}^{1}-\int_{-1}^{0}\right] d \cos \theta_{\ell}\frac{d^{2}
(\Gamma+\bar{\Gamma})}{d q^{2} d \cos \theta_{\ell}} }{d( \Gamma+\bar{\Gamma}) / d q^{2}}=\frac{3}{4}A_{6s}.
\label{eq:AFB}
\eeq

\item[(3)]
The $\phi$ polarization fractions:
\beq
F_L(q^2)= \frac{1}{4} (3S_{1c}-S_{2c}),\quad F_T(q^2)=\frac{1}{2}(3S_{1s}-S_{2s}).
\eeq
In the massless limit, since the CP-averaged observable $S_{1(c,s),2(c,s)}$ obey the relations $S_{1s}=3S_{2s}$ and $S_{1c}=-S_{2c}$,  the definitions of the polarization fractions can be simplified
directly as:
\beq
F_L(q^2)=S_{1c}=-S_{2c}, \quad  F_T(q^2)=\frac{4}{3}S_{1s}=4S_{2s}.
\eeq

\item[(4)]
The clean (no S-wave pollution) observables $P_{1,2,3}$ and $P^\prime_{4,5,6}$ in the natural basis  can be defined in terms of the coefficients  $S_i$ through the following relations
\cite{Kruger:2005ep,Becirevic:2011bp,Wang2015}:
\beq
P_{1} &=&  \frac{S_{3}}{2 S_{2s}}, \quad P_{2} =  \beta_\ell\frac{S_{6s}}{8 S_{2s}} ,\quad P_{3}  =  -\frac{S_{9}}{4 S_{2s}}\ , \\
P'_{4}  &=&  \frac{S_4}{\sqrt{S_{1c} S_{2s}}}, \quad P'_{5} =  \frac{\beta_\ell S_5}{2\sqrt{S_{1c} S_{2s}}} , \quad
P'_{6}  =- \frac{\beta_\ell S_7}{2\sqrt{S_{1c} S_{2s}}} ,\quad  P'_{8}  =- \frac{S_8}{\sqrt{S_{1c} S_{2s}}} ,  \label{eq:P series}
\eeq
where $\beta_\ell=\sqrt{1-4m_{\ell}^2/q^2}$.

\item[(5)]
In the massless limit of leptons,  the optimised observables $P^{(\prime)}_i$   \cite{Matias:2012xw} can be transformed as the following form:
\beq
P_{1} &=&  \frac{2S_{3}}{F_T} \quad P_{2} = \frac{S_{6s}}{2 F_T}, \quad  P_{3} = \frac{-S_{9}}{F_T},  \\
P'_{4} &=& \frac{2S_4}{\sqrt{F_L (1- F_L)}},\quad  P'_{5}  =  \frac{S_5}{\sqrt{F_L (1- F_L)}} ,\non
P'_{6} &= &-\frac{ S_7}{\sqrt{F_L (1- F_L)}},\quad P'_{8}  =- \frac{2S_8}{\sqrt{F_L (1- F_L)}} .
\eeq
\end{itemize}

One should know that our definitions of  the CP averaged angular coefficients $S_i$ , the CP asymmetry angular coefficients $A_i$ and the clean observable $P_{1,2,3}$ and$P^\prime_{4,5,6}$
differ from those adopted by the LHCb collaboration. To be specific, the reasons are the following:
\begin{itemize}
\item[(1)]
Our conventions for the angles to define the $B_s \to \phi \ell^+ \ell^-$ kinematics are identical to the Ref.~\cite{Altmannshofer:2008dz} but different from
the LHCb choices \cite{Aaij:2015oid,Aaij:2015esa}. The corresponding relations are the following :
\beq
\thK^{\rm LHCb}=\thK, \quad \thl^{\rm LHCb}=\pi-\thl,\quad \Phi^{\rm LHCb}=-\Phi.
\eeq
Some angular coefficients $I_i$ , $S_i$ and $A_i$, consequently, will have different signs:
\beq
I^{\rm LHCb}_{4,6,7,9}=-I_{4,6s,7,9},\quad S^{\rm LHCb}_{4,6,7,9}=-S_{4,6s,7,9},\quad A^{\rm LHCb}_{4,6,7,9}=-A_{4,6s,7,9}.
\eeq
Other remaining coefficients $I_i$ ($S_i$ and $A_i$), however, have the same sign in both conventions.

\item[(2)]
Our definitions of the clean observables $P_{1,2,3}$ and $P^\prime_{4,5,6,8}$ in Eq.~(\ref{eq:P series}) in terms of  $S_i$ may be different from those  defined and used by the LHCb Collaboration for example
in Ref.~\cite{Aaij:2015oid}. The resultant differences of the sign and normalization  are of the following:
\beq
P^{\rm LHCb}_{1}=P_1,\quad P^{\rm LHCb}_{2,3}=-P_{2,3}, \quad
P^{\prime \rm LHCb}_{4,8}=-\frac{1}{2}P^{\prime}_{4,8},\quad P^{\prime \rm LHCb}_{5,6}=P^{\prime}_{5,6}.
\eeq
\end{itemize}
For more details about the angular conventions of the angular observables of the semileptonic decays $B_{(s)} \to V l^+ l^-$, one can see Ref.~\cite{prd93:2016ghz}.


\section{Numerical results and discussions} \label{sec:4}

In the numerical calculations we use the following input parameters (here masses and decay constants are in units of GeV) \cite{pdg2018,Ali:2007ff}:
\beq
 \Lambda^{f=4}_{\overline{\rm MS}}&=&0.250, \; \quad \tau_{B_s^0}=1.509 {\rm ps},\; \quad m_b=4.8,\; \quad m_W=80.38, \;\quad  m_{\phi}=1.019 \non
 m_{B_s}&=&5.367, \; \quad m_{e}=0.000511, \; \quad m_{\mu}=0.105, \; \quad m_{\tau}=1.777, \non
 f_{B_s}&=&0.23, \quad f_\phi=0.231(4), \quad f^{\perp}_\phi=0.20(1), \quad a^{||}_{2\phi} =0.18(8), \quad a^{\perp}_{2\phi} =0.14(7).   \label{eq:inputs}
\eeq
For the CKM matrix elements and angles, we adopt the following values as given in Ref.~\cite{pdg2018}:
\beq
V_{tb}&=&1.019(25), \quad V_{us}= 0.2243(5),  \quad |V_{ts}|=(39.4\pm2.3)\times10^{-3}, \non
|V_{ub}|&=&(3.94\pm0.36)\times 10^{-3}, \quad 2\beta_s=0.021(31), \quad\gamma=\left ( {73.5}^{+4.2}_{-5.1} \right )^{\circ}.
\eeq

\subsection{The form factors }

For the considered semileptonic decays, the differential decay rates and other physical observables strongly rely on the value and the shape of
the relevant form factors $V(q^2),$ $A_{0,1,2}(q^2)$ and $T_{1,2,3}(q^2)$ for $B_s \to \phi \ell^+ \ell^-$ decays.
These form factors have been calculated in rather different theories or models \cite{Straub:2015ica,Keum:2000wi,Lu:2000em,li2003,Ali:2007ff}.
Since the PQCD predictions for the considered form factors are valid only at the large hadronic recoil ( low-$q^{\rm 2}$) region, we usually calculate explicitly
the values of the relevant form factors  at the low-$q^2$ region,  say  $ 0 \leq q^{\rm 2} \leq m_{\rm \tau}^{\rm 2}$,
and then make an extrapolation for all relevant form factors from the low-$q^{\rm 2}$ towards the high-$q^{\rm 2}$ region by using the
pole model parametrization\cite{cheng2004,wang2009} or other different methods.

In Refs.~\cite{Fan:2015,Hu:2019,Hu:2020}, we developed a new method:  the so-called  ``PQCD+Lattice"  approach.
Here we still use the PQCD approach to evaluate the form factors at the low $q^2$ region, but take those currently available lattice QCD results for the relevant
form factors at the high-$q^2$ region as the lattice QCD input to improve  the extrapolation  of  the form factors  up to  $q^2_{max}$.
In Refs.~\cite{Hu:2019,Hu:2020}, we used the Bourrely-Caprini-Lellouch (BCL) parametrization method \cite{bcl09,jhep1905-094} instead of the traditional
pole model parametrization.

\begin{table}[thb]
\caption{The  PQCD  predictions of the seven form factors $F_{B_s \to \phi}^i (0)$  with the theoretical uncertainties from the variations
of the parameters $\omega_{B_s}$, $a^{||,\perp}_{2\phi}$, $f_{\phi}$ and$f^{T}_{\phi}$.  } \label{tab:table2}
\centering
\setlength{\tabcolsep}{6pt} 
\renewcommand{\arraystretch}{1.5} 
\begin{tabular}{c|c }  \hline \hline
$F_{B_s \to \phi}^i (0)$ &  PQCD predictions  \\ \hline
$V(0)$    &     ${0.311}$ $^{+0.063}_{-0.051}(\omega_{B_s})$   $^{+0.008}_{-0.007}( a^{\perp}_{2\phi})$ $\pm 0.004 ( f_{\phi} )$   $\pm  0.004( f^{T}_{\phi} )$  \\ \hline
$A_0(0)$ &     ${0.262}$ $^{+0.046}_{-0.038}(\omega_{B_s})$   $^{+0.009}_{-0.008}( a^{||}_{2\phi})$ $\pm 0.002 ( f_{\phi} )$   $\pm  0.008( f^{T}_{\phi} )$  \\
$A_1(0)$ &     ${0.247}$ $^{+0.051}_{-0.041}(\omega_{B_s})$   $^{+0.005}_{-0.006}( a^{\perp}_{2\phi})$ $\pm 0.003 ( f_{\phi} )$   $\pm  0.003( f^{T}_{\phi} )$  \\
$A_2(0)$ &     ${0.239}$ $^{+0.054}_{-0.042}(\omega_{B_s})$   $\pm 0.009 ( a^{||,\perp}_{2\phi})$  $\pm 0.004 ( f_{\phi} )$   $\pm  0.001( f^{T}_{\phi} )$  \\ \hline
$T_{1,2}(0)$ &     ${0.264}$  $^{+0.052}_{-0.042}(\omega_{B_s})$   $\pm 0.006 ( a^{\perp}_{2\phi})$ $\pm 0.003 ( f_{\phi} )$   $\pm  0.005( f^{T}_{\phi} )$  \\
$T_3 (0)$ &     ${0.196}$      $^{+0.041}_{-0.033}(\omega_{B_s})$  $\pm 0.008 ( a^{||,\perp}_{2\phi})$  $\pm 0.003 ( f_{\phi} )$   $\pm  0.001( f^{T}_{\phi} )$  \\
\hline
\end{tabular}
\end{table}

\begin{table}[thb]
\caption{The central values of the theoretical predictions for the form factors of $B_s\to  \phi $ transitions at $q^2=0$ from  different theories or models
\cite{Ali:2007ff,Li:2009tx,Ball:2004rg,Yilmaz:2008pa,Straub:2015ica,Peng:2020ivn,Melikhov:2000yu,Faustov:2013pca,Lu:2007sg,Wu:2006rd,Su:2011eq,Dubnicka:2016nyy} .  }
\label{tab:table3}
\centering
\begin{tabular}{c|ccc|ccc} \hline\hline
     &  $V(0)$ & $A_0(0)$ &$A_1(0)$&$A_2(0)$ &$T_{1,2}(0)$&$T_3(0)$ \\
\hline
This work &$0.311$ &$0.262$  &$0.247$ & $0.239$& $0.264$& $0.196$ \\
PQCD\cite{Ali:2007ff} &$0.25$ &$0.30$&$0.19$ &$-$&$-$  \\
PQCD\cite{Li:2009tx} &0.26& $0.31$& $0.18$& $0.12$& $0.23$& $0.19$ \\\hline
LCSR\cite{Straub:2015ica}   &$0.387$&$0.389$&$0.296$& $-$&$0.309$&$-$\\
LCSR\cite{Ball:2004rg}   & $0.434 $& $0.474$& $0.311$& $0.234$& $0.349$&$0.175$\\
LCSR\cite{Yilmaz:2008pa} &$0.433$&$0.382$&$0.296$& $0.255$&$0.348$&$0.254$\\
QCDSR\cite{Peng:2020ivn}  & $0.45$& $0.30$& $0.32$& $0.30$& $0.33$&$0.22$\\
RDA\cite{Melikhov:2000yu} &$0.44$&$0.42$&$0.34$&$0.31$&$0.38$&$0.26$\\
RQM\cite{Faustov:2013pca} &$0.406$&$0.322$&$0.320$&$0.318$&$0.275$&$0.133$\\
SCET\cite{Lu:2007sg} &$0.329$&$0.279$&$0.232$&$0.210$&$0.276$&$0.170$\\
HQEFT\cite{Wu:2006rd} &$0.339$&$0.269$&$0.271$&$0.212$&$0.299$&$0.191$ \\
SQEH\cite{Su:2011eq} &$0.259$&$0.311$&$0.194$&$-$&$-$&$-$  \\
CQM\cite{Dubnicka:2016nyy}&$0.31$&$0.28$&$0.27$&$0.27$&$0.27$&$0.18$ \\ \hline\hline
\end{tabular}
\end{table}

In Table \ref{tab:table2}, we list the PQCD predictions for all seven relevant  form factors  $V(0)$, $A_{0,1,2}(0)$ and $T_{1,2,3}(0)$ for $B_s\to \phi $ transition at the point $q^2=0$.
The dominant theoretical errors come from the uncertainties of the parameter $\omega_{B_s}=0.50\pm 0.05$  \cite{Ali:2007ff} ,  the Gegenbauer moments
$a_{2\phi}^{||}=0.18\pm0.08$ and $a_{2\phi}^{\perp}=0.14\pm 0.07$  \cite{Ball:2006wn,Ali:2007ff,xiao18a,xiao18b},
as well as the decay constants $f_{\phi}=0.231\pm 0.004$ GeV and $f^{T}_{\phi}=0.20\pm 0.01$  GeV \cite{Ali:2007ff}.

In Table \ref{tab:table3}, as a comparison, we also list the central values of the theoretical predictions for the form factors    $F_{B_s \to \phi}^i (0)$  at $q^2=0$
evaluated in the PQCD approach \cite{Ali:2007ff,Li:2009tx},  and in other different theories or models
\cite{Ball:2004rg,Yilmaz:2008pa,Straub:2015ica,Peng:2020ivn,Melikhov:2000yu,Faustov:2013pca,Lu:2007sg,Wu:2006rd,Su:2011eq,Dubnicka:2016nyy}.
One can see that there exist  always some differences between different authors, even among the authors using the same approach.
Taking the calculations based on the LCSR method as an example, the authors of Ref.~\cite{Straub:2015ica} introduced the hadronic input parameters,  Ball and Zwicky
considered the one-loop radiative corrections \cite{Ball:2004rg}, Yilmaz included the radiative and higher twist corrections and SU(3) breaking effects \cite{Yilmaz:2008pa}.

In Table  \ref{tab:table4}, for the $B_s \to \phi$  transition form factors $(V,A_{0,1,2}, T_{1,2,3} )$, we quote directly the values of the lattice QCD results  at two or three reference points
of the high $q^2$ region, say $q^2=12,16$ GeV$^2$ and $q^2_{max}=(m_{B_s}-m_{\phi})^2\approx 18.9$ GeV$^2$,  as listed in Table XXXI of  Ref.~\cite{Horgan:2013hoa}.
In Ref.~\cite{Horgan:2013hoa}, the authors defined the helicity form factors $A_{12}(q^2)$ and $T_{23}(q^2)$ from the ordinary form factors
$A_{1,2}(q^2)$ and $T_{2,3}(q^2)$:
\beq
A_{12}(q^2)&=&\frac{(m_{B_s}+m_{\phi})^2(m^2_{B_s}-m^2_{\phi}-q^2)A_1(q^2)-\lambda(q^2) A_2(q^2)}{16 m_{B_s}m^2_{\phi}(m_{B_s}+m_{\phi})}, \non
T_{23}(q^2)&=&\frac{m_{B_s}+m_{\phi}}{8m_{B_s}m^2_{\phi}}\left[(m^2_{B_s}+3m^2_{\phi}-q^2)T_2(q^2)-\frac{\lambda(q^2) T_3(q^2)}{m^2_{B_s}-m^2_{\phi}} \right], \label{eq:a12t23}
\eeq
where the kinematic variable $\lambda(q^2)=(t_{+}-q^2)(t_{-}-q^2)$  with $t_{\pm}=(m_{B_s} \pm m_{\phi})^2$.
From above two equations and the numerical values of  $(A_1(q^2),T_2(q^2),A_{12}(q^2),T_{23}(q^2))$  as given in  Ref.~\cite{Horgan:2013hoa}, we can
find the corresponding lattice QCD results of $A_{2}(q^2)$ and $T_{3}(q^2)$ at the two points $q^2=(12,16)$ GeV$^2$ by direct numerical calculations.
When $q^2 \to q^2_{max}=( m_{B_s}-m_{\phi})^2$,  however, the parameter $\lambda(q^2_{max})$  in Eq.~(\ref{eq:a12t23})  is  also approaching zero simultaneously,
one therefore can not determine $A_{2}(q^2_{max})$ and $T_{3}(q^2_{max})$ reliably from  the values of $A_{12}(q^2_{max})$ and $T_{23}(q^2_{max})$
as given in Ref.~\cite{Horgan:2013hoa}. Consequently,  $A_{2}(q^2_{max})$ and $T_{3}(q^2_{max})$ are absent in Table \ref{tab:table4}.

\begin{table}[thb]
\caption{The values for the lattice QCD results of the relevant $B_s \to \phi $ transition form factors at two or three reference points ~\cite{Horgan:2013hoa}.  \label{tab:table4} }
\centering
\setlength{\tabcolsep}{6pt} 
\renewcommand{\arraystretch}{1.5} 
\begin{tabular}{c|c|ccc|ccc} \hline\hline
 $q^2$  & $V(q^2)$ & $A_0(q^2)$ & $A_1(q^2)$& $A_{2}(q^2)$ & $T_1(q^2)$& $T_2(q^2)$&$ T_{3}(q^2)$ \\  \hline
12 &$0.77(6)$  & $0.90(6)$  &$0.44(3)$  &$0.48(4)$ &$0.69(4)$  &$0.45(3)$ & $0.46(4)$\\
16 &$1.19(7)$  & $1.32(7)$  &$0.52(3)$  &$0.54(4)$ &$0.99(5)$  &$0.53(3)$ & $0.70(5)$\\
18.9 &$1.74(10)$  & $1.85(10)$  &$0.62(3)$  &$-$ &$1.36(8)$  &$0.62(3)$ & $-$\\ \hline \hline
\end{tabular}
\end{table}

\begin{figure}[thb]
\begin{center}
\centerline{\epsfxsize=5.5cm\epsffile{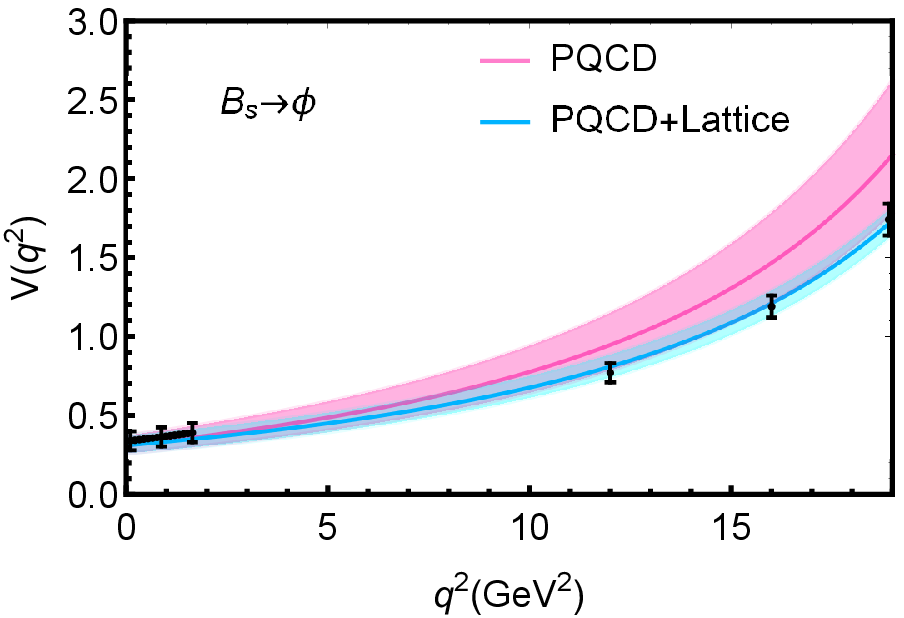} \vspace{0.3cm} \epsfxsize=5.5cm\epsffile{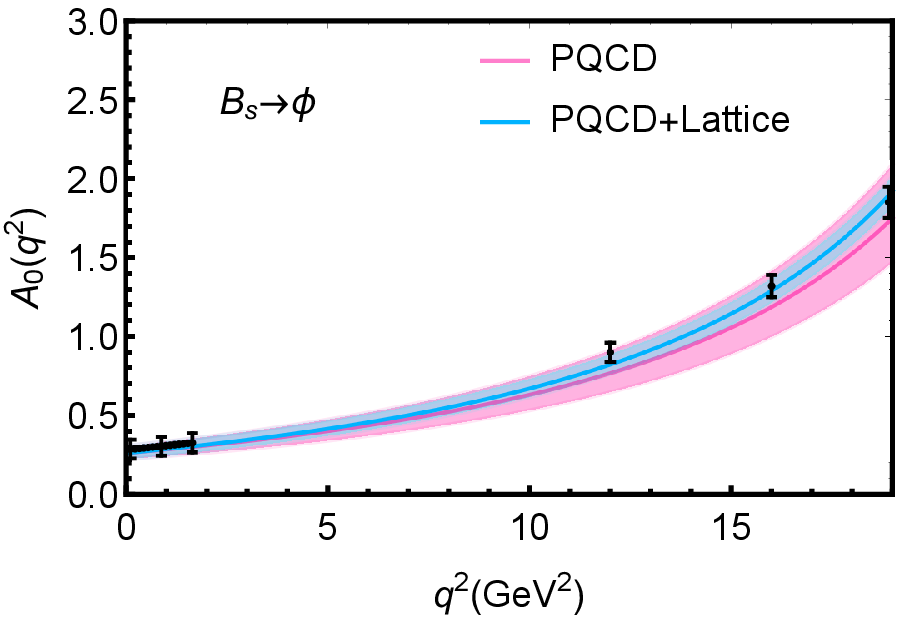} }\vspace{0.3cm}
\centerline{\epsfxsize=5.5cm\epsffile{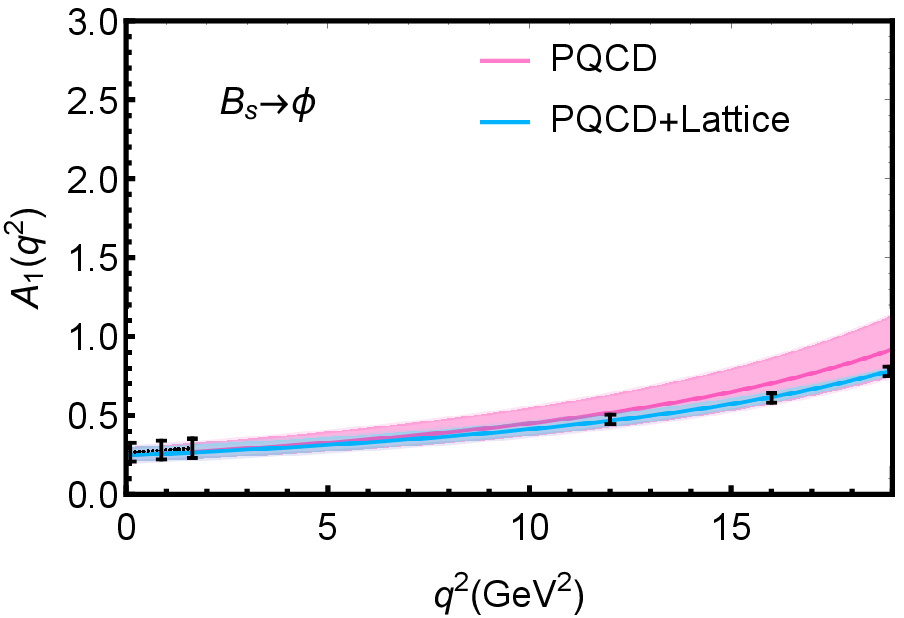} \vspace{0.3cm} \epsfxsize=5.5cm\epsffile{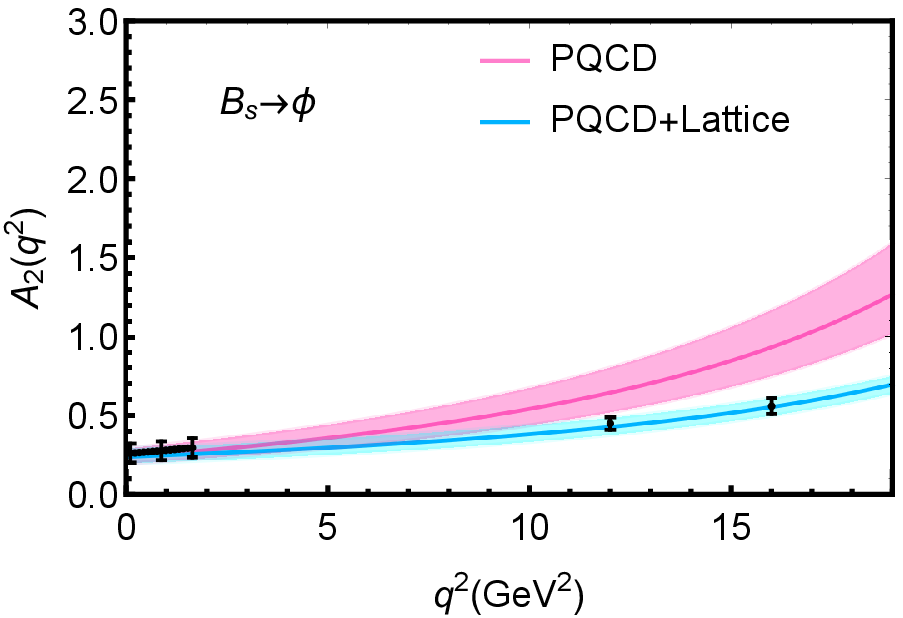}}\vspace{0.3cm}
\centerline{\epsfxsize=5.5cm\epsffile{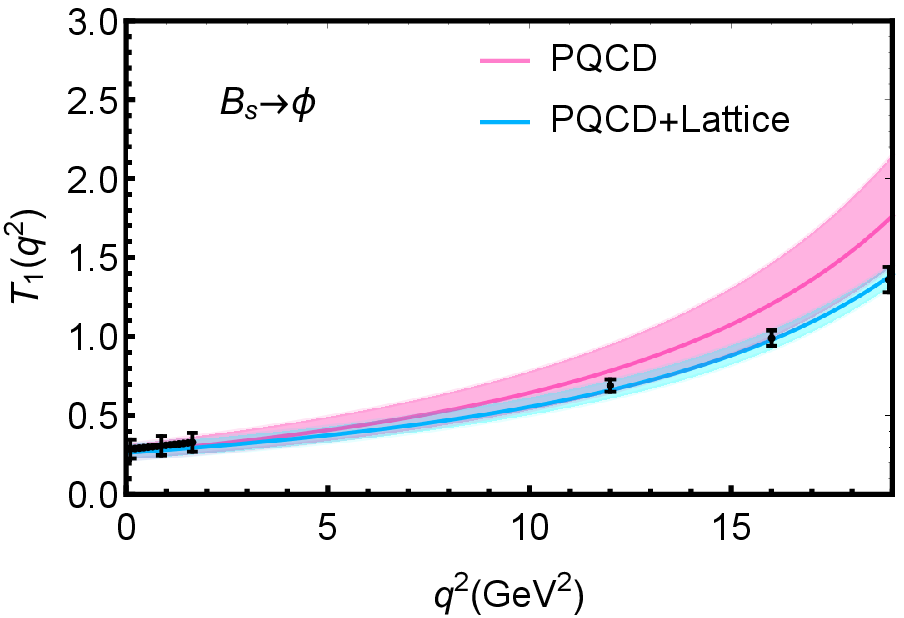} \vspace{0.3cm}\epsfxsize=5.5cm\epsffile{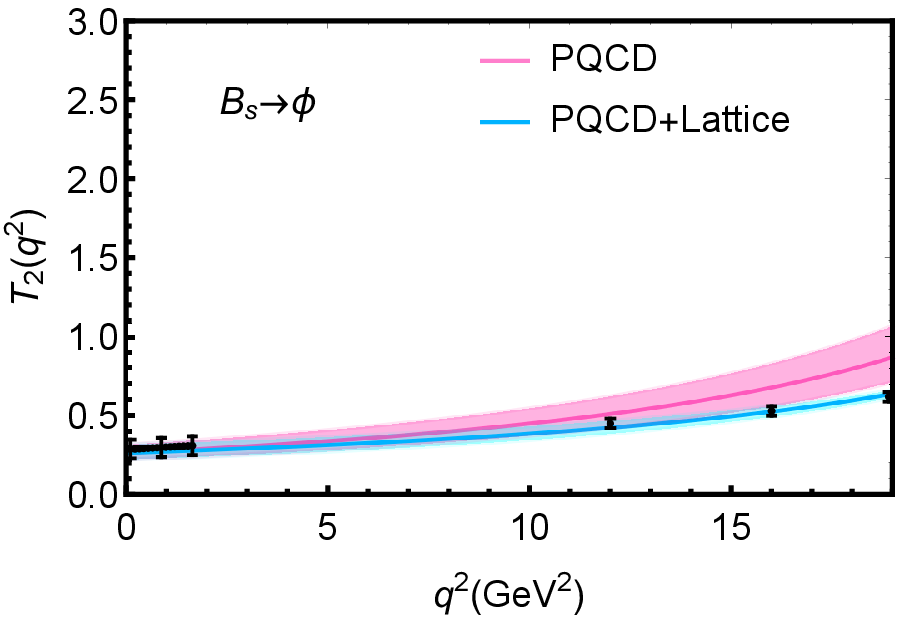}}\vspace{0.3cm}
\centerline{ \epsfxsize=5.5cm\epsffile{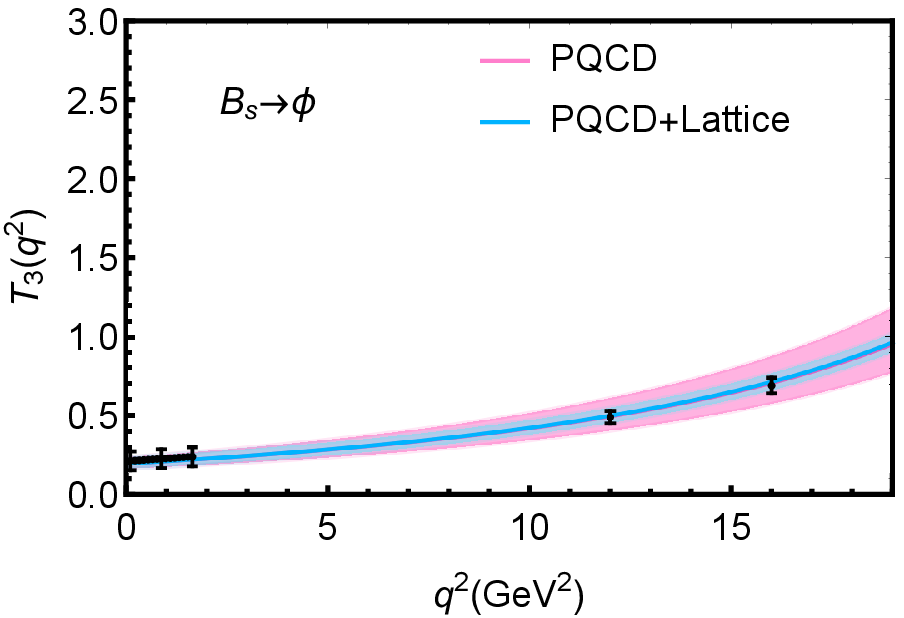} \vspace{0.3cm} \hspace{5.5cm} }
\end{center}
\vspace{-1cm}
\caption{Theoretical predictions of the relevant form factors for $B_s \to \phi$ transition in the PQCD approach  ( red curves) and the  ``PQCD+Lattice"  approach (blue curves).
The red (blue) shaded band represents the theory uncertainties.  The black error bars in the low-$q^2$ region correspond to  the PQCD predictions.
The error bars in high-$q^2$ region are the lattice QCD results. }
\label{fig:fig1}
\end{figure}

In this work, we will use both the PQCD factorization approach and the ``PQCD+Lattice" approach to evaluate all relevant form factors over the whole  range of $q^2$.
\begin{enumerate}
\item[(1)]
In the PQCD approach,  we use the definitions and formulae to calculate the values of  all relevant form factors $V(q^2)$ , $A_{0,1,2}(q^2)$ and $T_{1,2,3}(q^2)$ at some points in the region of
$0 \leq q^2 \leq m_\tau^2$. We then make the extrapolation for these form factors  to the  large $q^2$ region up to $q_{max}^2 $ by using the selected parametrization method directly.

\item[(2)]
In the ``PQCD+Lattice " approach,  we  take  the lattice QCD results of  the form factors at the points  of $q^2=(12,16,18.9)$ GeV$^2$ as the new input in the high $q^2$ region
and then  make a combined fit of the PQCD results in low $q^2$ region and the lattice QCD results in high $q^2$ region to determine the relevant parameters $b_k^i$
in the $z$-series expansion and then complete the extrapolation.

\item[(3)]
For both approaches,  we always use the model-independent $z$-series parametrization, which is based on a rapidly converging series
in the parameter $z$, as in Refs.~\cite{Kindra:2018ayz,Straub:2015ica} to make the extrapolation. The entire cut $q^2$-plane will be mapped
onto the unit disc $|z(q^2)|\leq1$ under the conformal transformation as \cite{Lu:2018cfc}
\beq
z(q^2)=\frac{\sqrt{t_{+}-q^2}-\sqrt{t_{+}-t_0}}{\sqrt{t_{+}-q^2}+\sqrt{t_{+}-t_0}}
\eeq
where $t_{\pm}=(m_{B_s}\pm m_{\phi})^2$ and $0\leq t_0 < t_{-}$  is a auxiliary parameter which can be optimised to reduce the maximum value
of $|z(q^2)|$ in the physical range of the form factors and will be taken in the same way as in Ref.~\cite{Bharucha:2010im}: $ t_0=t_{+}(1-\sqrt{1-t_{-}/t_{+}})$.
The form factors are finally parameterized in the BCL  version of the $z$-series expansion \cite{bcl09}
\beq
F_{B_s \to \phi}^{i}(q^2) &=& \frac{ F_{B_s \to \phi}^{i}(0)}{1-q^2/m_{i, \, \rm pole}^2} \, \left \{ 1 + \sum_{k=1}^N \, b_k^i \, \left [ z(q^2, t_0)^k - z(0, t_0)^k \right ]  \right \}  \non
&=&  \frac{F_{B_s \to \phi}^{i}(0)}{1-q^2/m_{i, \, \rm pole}^2} \, \left \{1 + b_1^i \, \left [z(q^2, t_0) - z(0, t_0) \right ]  \right \} + \cdots .
\label{eq:z-series}
\eeq
Since the term $ |z(q^2,t_0)|^2 \le 0.04$ in the whole considered $q^2$ region,  the high order $N\geq 2$ terms in  Eq.~(\ref{eq:z-series})   should be  very small in magnitude and therefore can be neglected.
After the truncation at $N=1$, the  coefficient $b^i_1$  for the corresponding form factor  $F_{B_s \to \phi}^{i}(q^2) $  can be determined by fitting to the PQCD predictions at low $q^2$
region and the lattice QCD results in the high $q^2$ region.
Taking the form factor $A_2(q^2)$ as an example,   we calculate $A_2(q^2)$ first by employing the PQCD approach in the sixteen points in the low $0 \leq q^2 \leq m_\tau^2$ region,
take the lattice QCD results $A_2(12)=0.48\pm 0.04$ and $A_2(16)=0.54\pm 0.04$  as additional input,  and finally make the fitting for the parameter $b_1^{A_2}$ and find that:
\beq
b_1^{A_2}=-1.820\pm 0.148 {\rm (standard \ \ error)},
\eeq
with the goodness-of-fit $R^2=0.9978$.  For other form factors  $F^{i}(q^2)$ we find the results by following the same kind of procedure.
The input values of the various $\bar{s}b$-resonance mass $m_{i,pole}$ in Eq.~(\ref{eq:z-series}) can be found from Ref.~\cite{pdg2018} and
are collected in Table \ref{tab:table5}. For further discussions on the systematic uncertainties due to the dependence of truncation schemes and on the implementation of the strong unitary constraints
one can see  Refs.~\cite{Bharucha:2010im,Bigi:2016mdz}.

\begin{table}[thb]
\caption{The masses $m_{i,{\rm pole}}$ in Eq.~(\ref{eq:z-series}) for the form factors $F^i_{B_s\to \phi}(q^2)$ \cite{Straub:2015ica}. }
\label{tab:table5}
\begin{tabular}{l|l|l}   \hline  \hline
\setlength{\tabcolsep}{6pt} 
\renewcommand{\arraystretch}{1} %
$F_{B_s \to \phi}^{i}(q^2)$\ \   & $\bar{s}b (J^P)$  & $m_{i,{\rm pole}}$({\rm GeV})   \\ \hline
    ${A}_0 (q^2)$ & $B_{s}(0^{-})$ &  5.366 \\
 ${V}(q^2)$,  ${T}_1 (q^2)$  & $B_s^\ast(1^-)$ &  5.415 \\
 ${A}_{1,2} (q^2)$,\, ${T}_{2,3}(q^2)$ & $B_{s1}(1^+)$  & 5.829 \\   \hline  \hline
\end{tabular}
\end{table}

\item[(4)]
In Fig.~\ref{fig:fig1}, we show the theoretical predictions of the form factors $V(q^2)$,$A_{0,1,2}(q^2)$ and
$T_{1,2,3}(q^2)$ for $B_s \to \phi$ transition based on the PQCD approach  ( red curves) and the  ``PQCD+Lattice" approach (blue curves) with the extrapolation  from  $q^2=0$ to $q^2_{max}=(m_{B_s} -m_\phi)^2$
by applying the $z$-series parameterizations.
The shaded bands represent the total theoretical error  obtained by adding in quadrature of  the separate errors from the uncertainty of the parameter $\omega_{B_s}$, $a^{||,\perp}_{2\phi}$, $f_\phi$ and $f_\phi^{\perp}$.
The black error bars in the low-$q^2$ region correspond to  the PQCD predictions of the corresponding form factors,  while the error bars in the high-$q^2$ region
are the currently known lattice QCD results as collected in Table \ref{tab:table4}

\end{enumerate}

\subsection{Numerical  results }

We now proceed to explore the phenomenological aspects of the cascade decays $B_s \to \phi (\to K^-K^+) \ell^+ \ell^-$, which allow us  to define and compute a number of
physical observables and compare them with those measured by experiments. We first compare our results for the branching ratios and angular observables with
the experimental data reported by the LHCb Collaboration\cite{Aaij:2015esa,Aaij:2021pkz}.
As studied systematically in last section, the physical observables accessible in the semileptonic decays $B_s \to \phi \ell^+ \ell^-$
are the CP averaged differential branching fraction $d{\cal B}/dq^2$ \cite{Aaij:2015esa,Aaij:2021pkz}, the CP-averaged $\phi$ meson longitudinal polarization fraction $F_L$, the forward-backward asymmetry $\cala_{FB}$,
the angular coefficients $S_{i}$ and $A_i$, and the optimized observables $P_i$ and $P_j^\prime$ \cite{Descotes-Genon:2015uva}.
The CP asymmetry angular coefficients $A_{5,6,8,9}$ in the SM are induced by the weak phase from the CKM matrix.
For the $b \to s$ transition, the CP asymmetries proportional to ${\rm Im}(V_{ub}V^\ast_{us}/V_{tb}V^\ast_{ts})$, which is of order $10^{-2}$ \cite{Bobeth:2008ij} as measured by the LHCb
Collaboration (see Table 3 in Ref.~\cite{Aaij:2015esa}), but the statistical uncertainties are still large.
For these reasons, we will focus on the CP averaged quantities when taking the binned observables into consideration.

\begin{figure}[tb]
\begin{center}
\centerline{\epsfxsize=7.5cm\epsffile{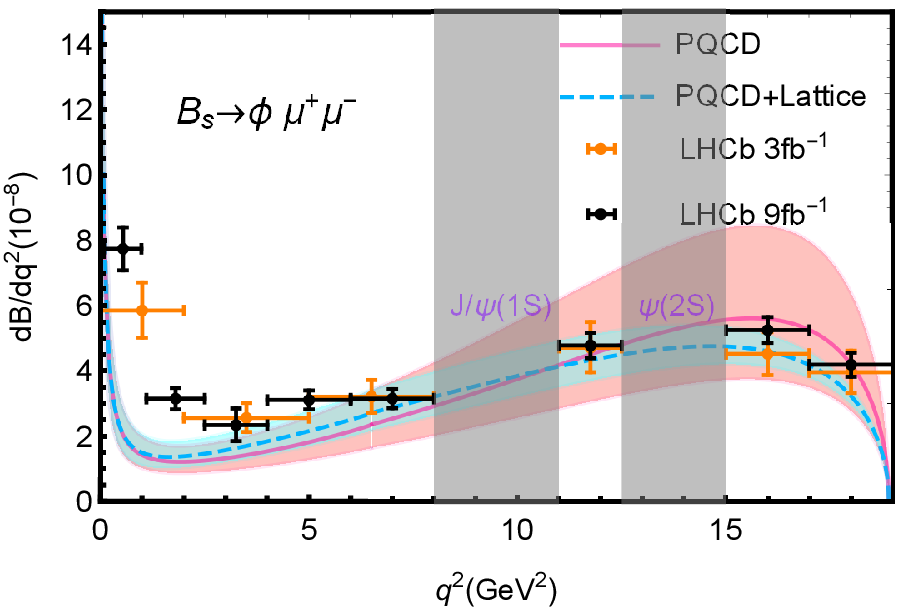} \hspace{0.3cm}\epsfxsize=7.5cm\epsffile{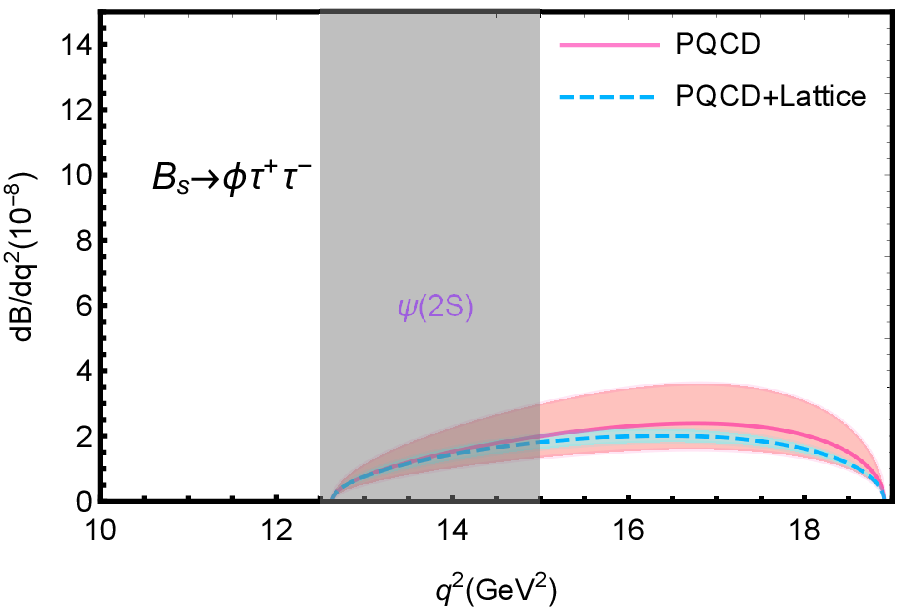} }
\end{center}
\vspace{-1cm}
\caption{Theoretical predictions for the $q^2$-dependence of differential branching fraction $d\calb/dq^2$ for the semileptonic decays $B_s \to \phi \mu^+ \mu^-$
in the PQCD (red band) and ``PQCD+Lattice" (blue band) approach, respectively. The crosses show the LHCb measurements in different bins as given in Ref.~\cite{Aaij:2015esa} ( orange ones )
and in  Ref.~\cite{Aaij:2021pkz} (black ones). The vertical grey blocks are the experimental veto regions.}
\label{fig:fig2}
\end{figure}

We begin with the branching ratios of the decays  $B_s \to \phi \ell^+ \ell^-$. From the differential decay rates as defined in Eq.~(\ref{eq:dGdq2}),  it is straightforward to
make the integration over the range of $4m_\ell^2\!\leq\! q^2\!\leq\!(m_{B_s}-m_{\phi})^2$.
In order to be consistent with the choices made by LHCb Collaboration in their data analysis,
we here also cut off  the regions of dilepton-mass squared around the charmonium resonances
$J/\psi(1S)$ and $\psi(2S)$:  i.e., $8.0\! < \!q^2\! <\! 11.0\, GeV^2$ and $12.5\! <\! q^2\! <\! 15.0 \,GeV^2$  for $\ell=(e,\mu,\tau)$ cases.
We display the PQCD and ``PQCD+Lattice" predictions for the differential branching ratios  $d{\cal B}/dq^2$ in Fig.~\ref{fig:fig2} for the cases of $l=(\mu,\tau)$,
including currently available LHCb results in six or eight bins of $q^2$ \cite{Aaij:2015esa,Aaij:2021pkz} indicated by the crosses for $B_s \to \phi \mu^+ \mu^-$ decay.
From Fig.~\ref{fig:fig2}  one can  see that both PQCD and ``PQCD+Lattice" predictions for the differential branching ratios do agree well with the LHCb results within
the still large errors. Since the theoretical prediction  for the differential branching ratio of the electron mode is almost identical with the one of the  muon,
we do not draw the figure of $d{\cal B}(B_s \to \phi e^+ e^-)/dq^2$   in Fig.~\ref{fig:fig2}.

\begin{table}[tb]
\caption{Theoretical predictions for the total branching fractions ${\cal B}(B_s \to \phi \ell^+ \ell^-)$ ( in units of $10^{-7}$ )  in the PQCD (the first row) and ``PQCD+Lattice" (the second row) approaches.
As a comparison, we also list the LHCb measured value for muon channel corresponding to an integrated luminosity of 3$\rm fb^{-1}$\cite{Aaij:2015esa} and  9$\rm fb^{-1}$\cite{Aaij:2021pkz}  and the
QCDSR predictions for all three channels \cite{Peng:2020ivn}. }
\label{tab:table6}
\centering
\begin{tabular}{l | c | c | c }  \hline \hline
 BFs   & PQCD / ``PQCD+Lattice" & QCDSR  \cite{Peng:2020ivn}  &  LHCb  \\ \hline
$ {\cal B}(B_s \to \phi  e^+ e^-)$             & ${8.55} ^{+4.02}_{-2.69}(\rm FFs) \pm {0.15}(\mu) \pm{0.42}(V_{tb})  \pm{0.65}(V_{ts})$  & $ 7.12\pm 1.40 $&  \\
                                                               & ${8.24}  ^{+2.03}_{-1.53} (\rm FFs) \pm{0.14}(\mu) \pm{0.41} (V_{tb})  \pm{0.63} (V_{ts})$  &   &\\
$ {\cal B}(B_s \to \phi  \mu^+ \mu^-)$    & ${7.07} ^{+3.37}_{-2.25}(\rm FFs) \pm {0.12}(\mu) \pm{0.38}(V_{tb})  \pm{0.53}(V_{ts})$  &$7.06\pm 1.59$ &$ 7.97^{+0.81}_{-0.80}$\cite{Aaij:2015esa}\\
                                                               & ${6.76}  ^{+1.39}_{-1.09} (\rm FFs) \pm{0.11}(\mu) \pm{0.33} (V_{tb})  \pm{0.52} (V_{ts})$  & &$8.14\pm0.47$\cite{Aaij:2021pkz} \\
$ {\cal B}(B_s \to \phi  \tau^+ \tau^-)$    & ${0.81} ^{+0.42}_{-0.27}(\rm FFs) \pm {0.02}(\mu) \pm{0.04}(V_{tb})  \pm{0.06}(V_{ts})$  &$0.35\pm 0.17$ & \\
                                                               & ${0.68} ^{+0.06}_{-0.06}  (\rm FFs)         \pm {0.02}(\mu) \pm{0.03} (V_{tb})  \pm{0.05} (V_{ts})$  && \\ \hline\hline
\end{tabular}
\end{table}

In Table \ref{tab:table6} we present  the theoretical predictions of the total branching fractions for $B_s \to \phi \ell^+ \ell^-$ with $\ell=(e,\mu,\tau)$ obtained by the integration over the six $q^2$ bins
using the PQCD ( the first row) and ``PQCD+Lattice"  approach ( the second row), respectively.
The major theoretical errors from different sources, such as the form factors (FFs) as listed in Table \ref{tab:table2}, the scale $\mu$, the CKM matrix
element $V_{tb}$ and $V_{ts}$,  are also listed. As in Ref.~\cite{Aaij:2015esa}, a correction factor $f_{veto}=1.52$ is applied to account for the contribution in the veto $q^2$ bins for $\ell=(e,\mu)$ cases.
As a comparison, we also show the LHCb measured value ${\cal B}(B_s \to \phi \mu^+ \mu^-)=(7.97^{+0.81}_{-0.80}) \times 10^{-7}$  \cite{Aaij:2015esa}and $(8.14^{+0.47}_{-0.47})\times 10^{-7}$\cite{Aaij:2021pkz} and the QCDSR predictions
${\cal B}(B_s \to \phi \ell^+ \ell^-)$ for all three decay modes \cite{Peng:2020ivn}.
For $B_s \to \phi \mu^+ \mu^-$ decay, for instance,  the theoretical predictions and the LHCb measurement  \cite{Aaij:2015esa,Aaij:2021pkz}   (in unit of $10^{-7}$) are the following:
\beq
{\cal B}( B_s \to \phi \mu^+ \mu^-) =  \left \{ \begin{array}{ll}
7.07^{+3.43}_{-2.34},  & {\rm in \ \ PQCD },  \\
6.76^{+1.52}_{-1.25}, & {\rm in \ \ PQCD+Lattice },\\
7.06^{+1.59}_{-1.59},  & {\rm in \ \ QCDSR } \ [87],\\
7.97^{+0.81}_{-0.80}, & {\rm LHCb } \ \  [31]. \\
8.14^{+0.47}_{-0.47}, & {\rm LHCb } \ \  [33]. \\ \end{array} \right.
\label{eq:brtot1}
\eeq
From the numerical results in  above equation and  Table  \ref{tab:table6},  one can see that
\begin{enumerate}
\item[(1)]
The  PQCD and ``PQCD+Lattice"   predictions for the branching ratio  $\calb(B_s \to \phi \ell^+ \ell^-)$  with  $\ell=( e,\mu,\tau)$
do agree well  with each other within the errors, while the  ``PQCD+ Lattice" predictions of  $\calb(B_s \to \phi \ell^+ \ell^-)$
have  smaller errors than  those of  the PQCD predictions.

\item[(2)]
Both PQCD and ``PQCD+Lattice" predictions of  $\calb(B_s \to \phi \mu^+ \mu^-)$  do agree well with currently available  LHCb measured values  \cite{Aaij:2015esa,Aaij:2021pkz}
within errors. For the electron and tau mode, however, we have to wait for the future experimental measurements.

\item[(3)]
For all three decay modes, our theoretical predictions of the branching ratios do agree well
with the theoretical predictions obtained from the QCD sum rule  \cite{Peng:2020ivn}.
\end{enumerate}

Since the  large theoretical uncertainties of the branching ratios could  be largely canceled in the ratio of the branching ratios of $B_s\to \phi\ell^+\ell^-$ decays,
one can define and check the physical observables $R^{e\mu }_{\phi}$ and $R^{\mu \tau}_{\phi}$  \cite{Hiller:2003js}.
In the region $q^2< 4m^2_{\mu}$,  where only the $e^+ e^-$ modes are allowed, there is a large enhancement due to
the $1/q^2$ scaling of the photon penguin contribution \cite{Aubert:2008ps}.
In order to  remove the phase space effects in the ratio   $R^{e\mu }_{\phi}$ and keep consistent with other analysis \cite{Hiller:2003js},
we here also use the lower cut of $4 m^2_{\mu}$ for both the electron and muon modes in the definition of the ratio  $R^{e\mu }_{\phi}$ as in Ref.~\cite{Hiller:2003js}:
 \beq
R^{e\mu }_{\phi}=\frac{\int^{q^2_{max}}_{4m^2_{\mu}}dq^2\frac{d\calb(B_s\to \phi \mu^+\mu^-)}{dq^2} }{\int^{q^2_{max}}_{4m^2_{\mu}}dq^2
\frac{d\calb(B_s\to\phi e^+e^-)}{dq^2}  } =  \left\{\begin{array}{ll}
0.992\pm 0.002,  & {\rm in \ \ PQCD},\\
0.991 \pm 0.002, & {\rm in \ \ PQCD+Lattice}.\\  \end{array} \right.
\eeq
For the case of  the ratio  $R^{\mu\tau }_{\phi}$  we have
\beq
R^{\mu\tau }_{\phi}=\frac{\int^{q^2_{max}}_{4m^2_{\tau}}dq^2\frac{d\calb(B_s\to \phi \tau^+\tau^-)}{dq^2} }{\int^{q^2_{max}}_{4m^2_{\mu}}dq^2
\frac{d\calb(B_s\to\phi \mu^+\mu^-)}{dq^2}  } =  \left\{\begin{array}{ll}
0.115\pm 0.004,  & {\rm in \ \ PQCD},\\
0.100 \pm 0.009, & {\rm in \ \ PQCD+Lattice},\\  \end{array} \right.
\eeq
where the total error is the combination of the individual errors in quadrature. We suggest the LHCb and Belle-II to measure these two ratios.

\begin{figure}[thb]
\begin{center}
\centerline{\epsfxsize=6cm\epsffile{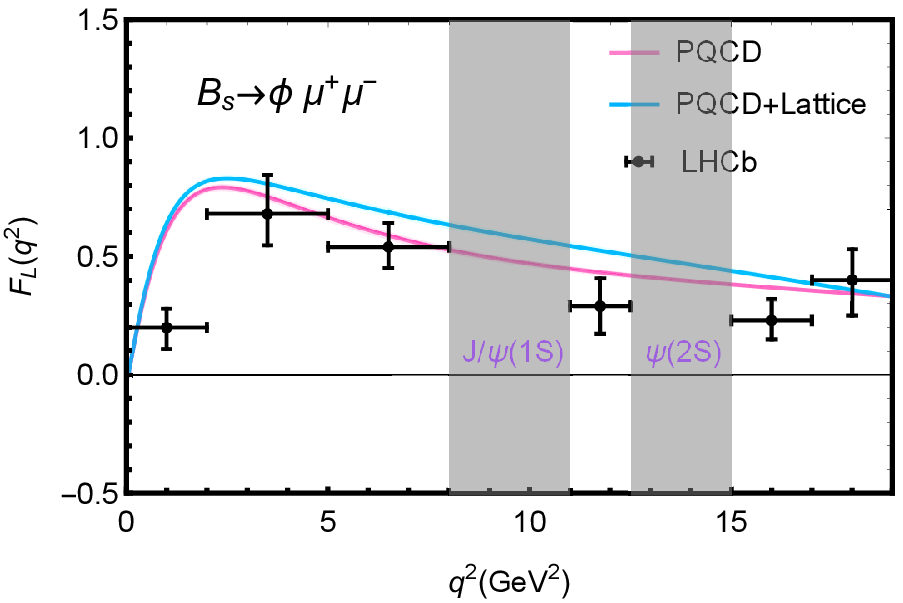} \hspace{0.5cm}\epsfxsize=6cm\epsffile{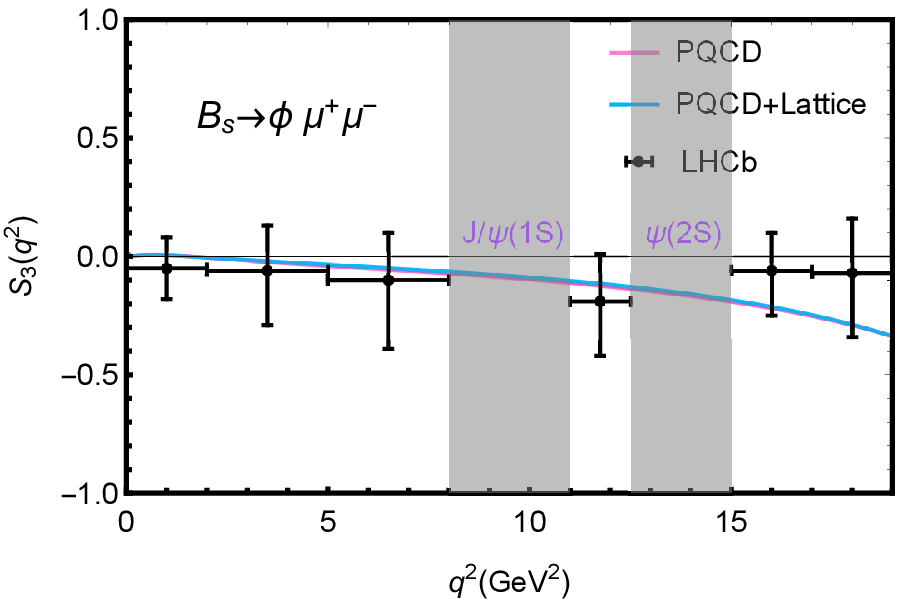} }\vspace{0.3cm}
\centerline{\epsfxsize=6cm\epsffile{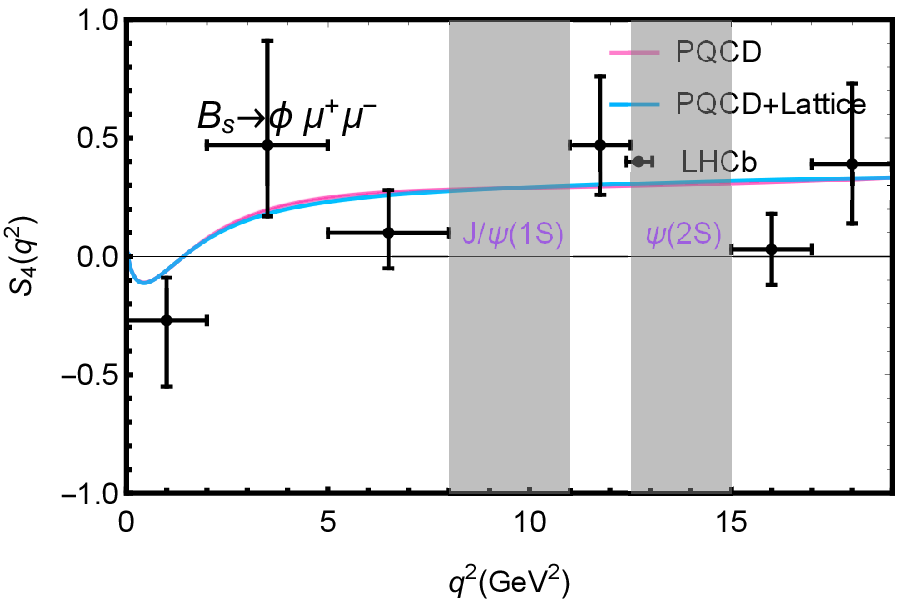}  \hspace{0.5cm} \epsfxsize=6cm\epsffile{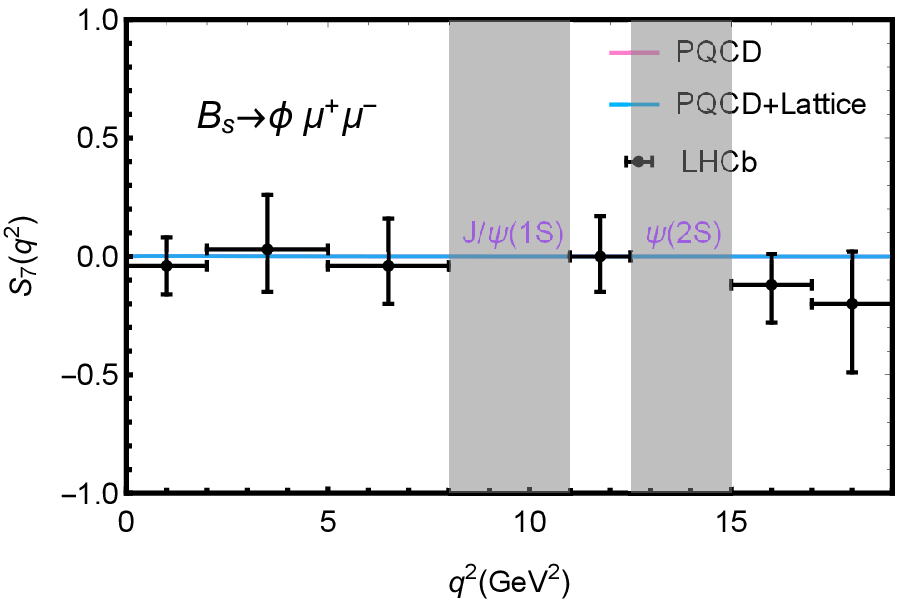} }\vspace{0.3cm}
\centerline{\epsfxsize=6cm\epsffile{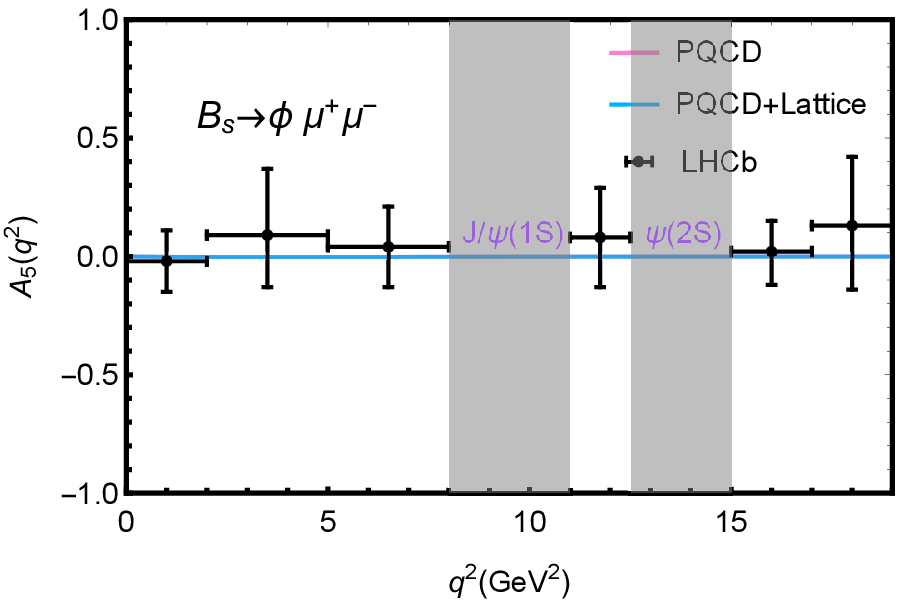}  \hspace{0.5cm} \epsfxsize=6cm\epsffile{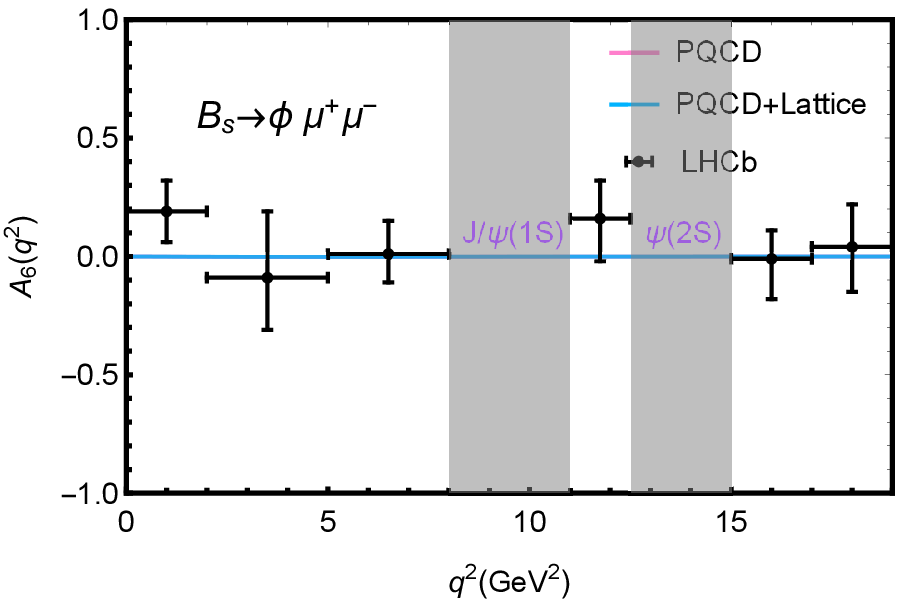} }\vspace{0.3cm}
\centerline{\epsfxsize=6cm\epsffile{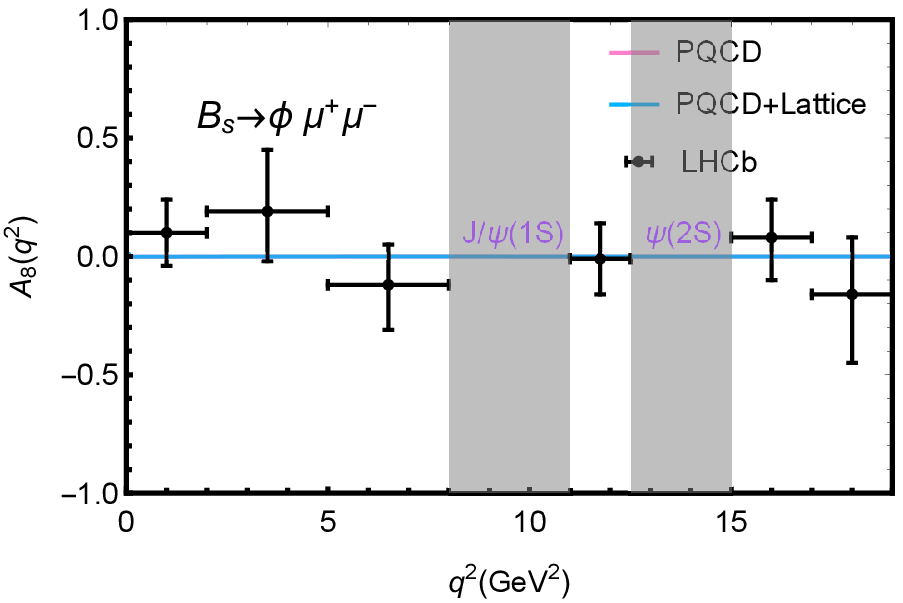}  \hspace{0.5cm} \epsfxsize=6cm\epsffile{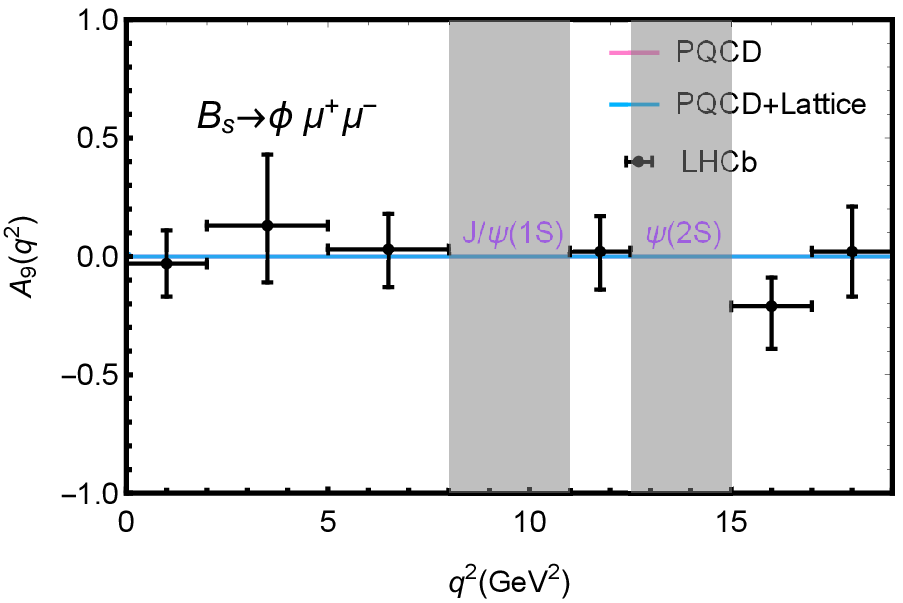} }\end{center}
\vspace{-1cm}
\caption{ Theoretical predictions for the $q^2$-dependence of the observables $F_{L}(q^2)$, $S_{3,4,7}(q^2)$ and $A_{5,6,8,9}(q^2)$
for the decay $B_s \to \phi \mu^+ \mu^-$  in the PQCD ( red lines ) and ``PQCD+Lattice" ( blue lines ) approach.
The crosses in each figure represent the LHCb measurements in six $q^2$ bins \cite{Aaij:2015esa}. The vertical grey blocks are the two experimental veto regions.}
\label{fig:fig3} \end{figure}

\begin{figure}[thb]
\begin{center}
\centerline{\epsfxsize=6cm\epsffile{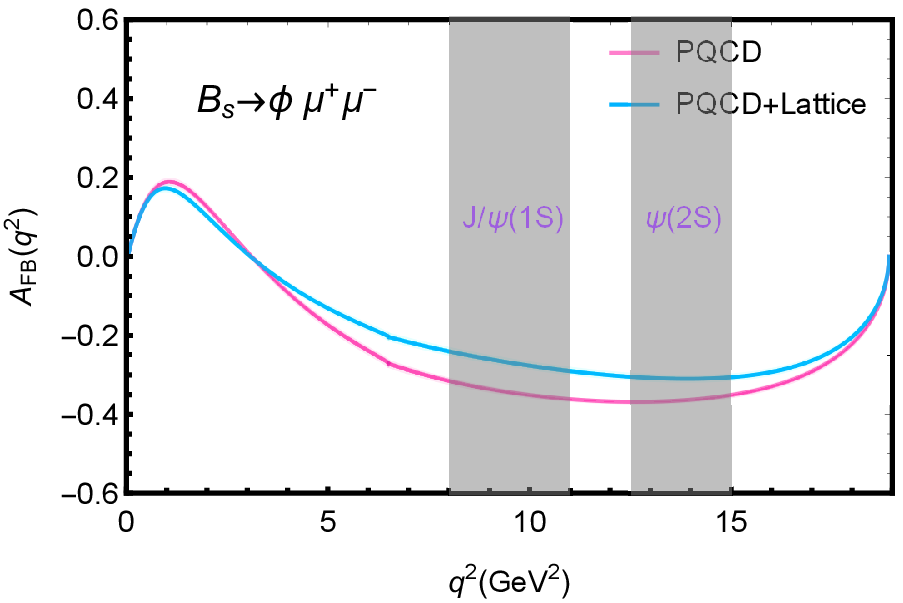}      \hspace{0.5cm} \epsfxsize=6cm\epsffile{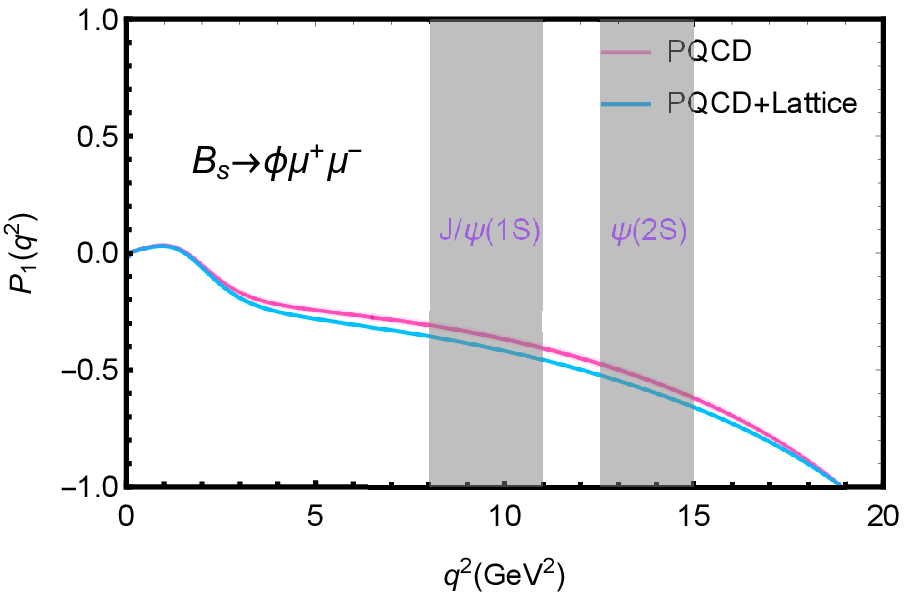} }\vspace{0.3cm}
\centerline{\epsfxsize=6cm\epsffile{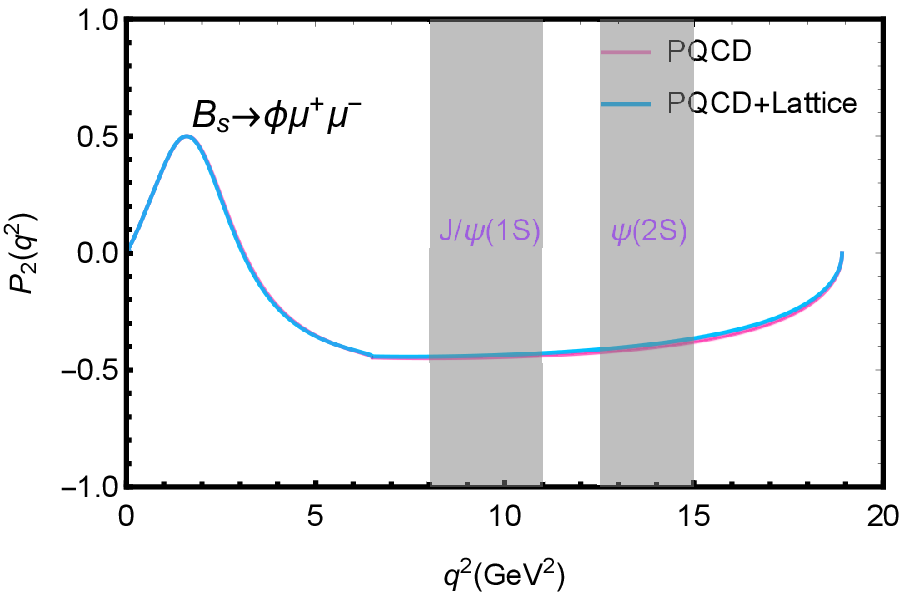}      \hspace{0.5cm}  \epsfxsize=6cm\epsffile{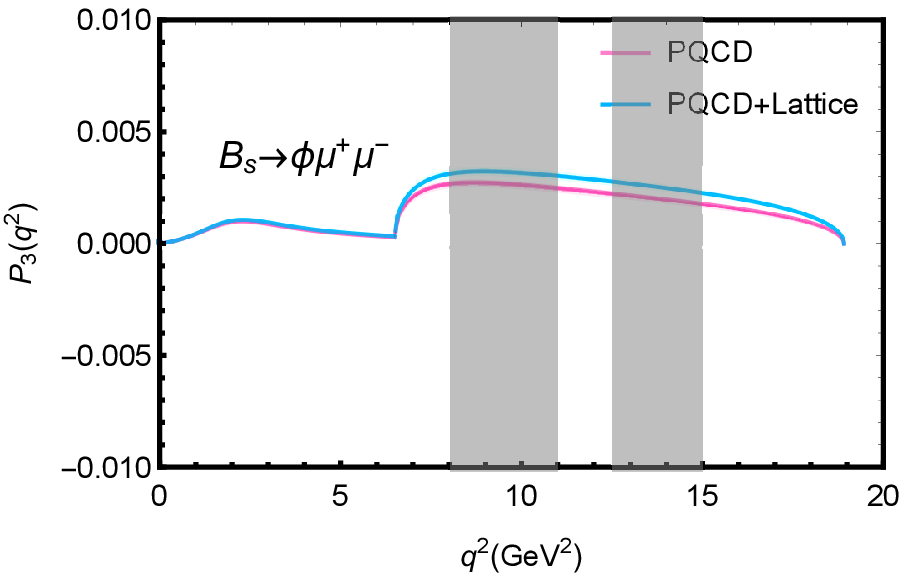} }\vspace{0.3cm}
\centerline{\epsfxsize=6cm\epsffile{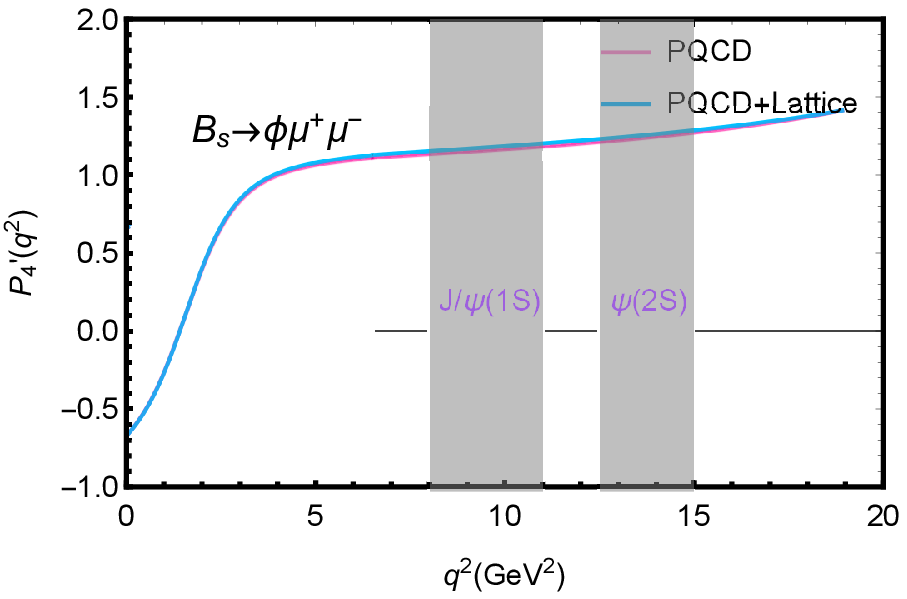}      \hspace{0.5cm} \epsfxsize=6cm\epsffile{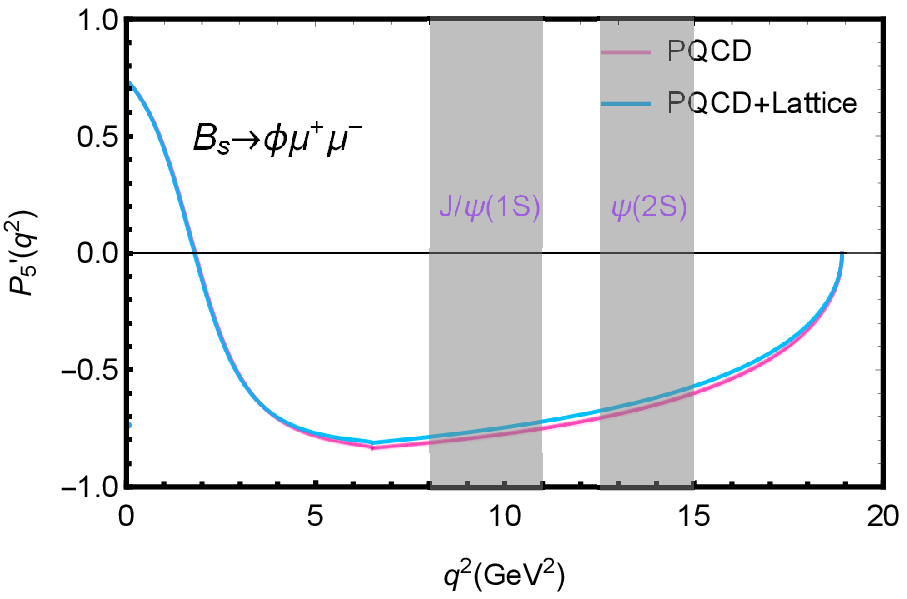} }\vspace{0.3cm}
\centerline{\epsfxsize=6cm\epsffile{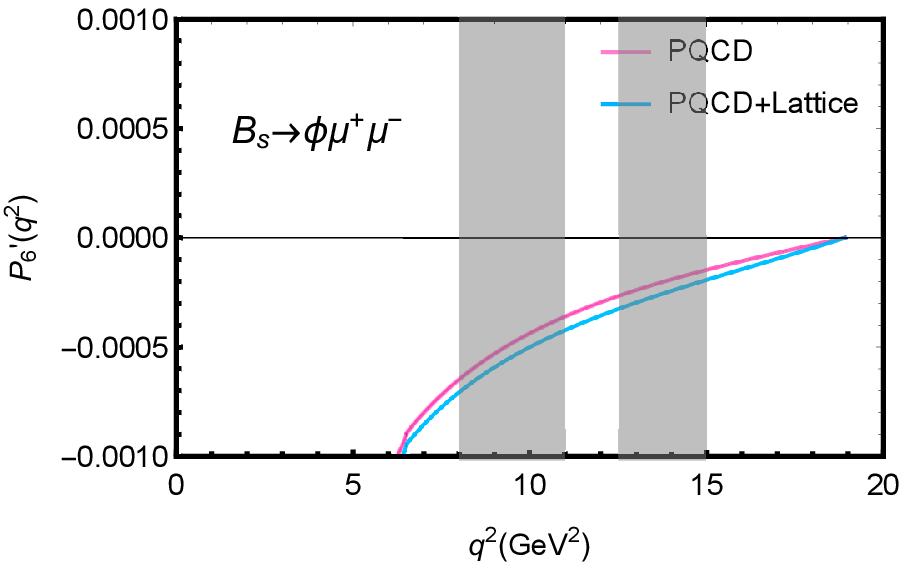}      \hspace{0.5cm} \epsfxsize=6cm\epsffile{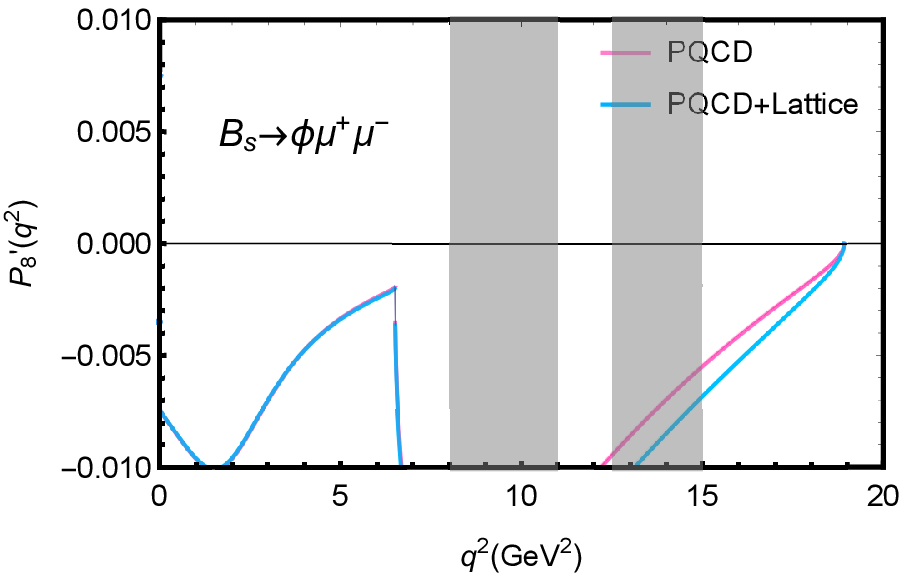} }
\end{center}
\vspace{-1cm}
\caption{Theoretical predictions for the $q^2$-dependence of the observables $A_{FB}(q^2)$,  $P_{1,2,3}(q^2)$ and $P^{\prime}_{4,5,6,8}(q^2)$
for $B_s \to \phi \mu^+ \mu^-$  decay in the PQCD (red lines) and ``PQCD+Lattice" (blue lines )approach.
The vertical grey blocks are the two experimental veto regions.}
\label{fig:fig4} \end{figure}

For $B_s \to \phi  \mu^+\mu^-$ decay,  we show our  theoretical predictions for the $q^2$-dependence of the longitudinal polarization
$F_{L}(q^2)$,  the CP averaged angular coefficients $S_{3,4,7}(q^2)$  and  the  CP asymmetry angular coefficients $A_{5,6,8,9}(q^2)$  in Fig.~\ref{fig:fig3}.
As a comparison,   the currently available LHCb measurements for  these observables of $B_s \to \phi \mu^+\mu^-$ decay in the six $q^2$ bins  \cite{Aaij:2015esa}
are also shown by those crosses explicitly. One can see from the Fig.~\ref{fig:fig3} that:
\begin{enumerate}
\item[(1)]
For the longitudinal polarization $F_{L}(q^2)$,  although both PQCD and ``PQCD + Lattice" predictions all agree well with the LHCb measurements in the six bins,
our theoretical predictions in the region of the fourth and fifth bin are little larger than the measured ones.

\item[(2)]
For  the CP averaged angular coefficients $S_{3,4,7}(q^2)$, the PQCD and ``PQCD + Lattice" predictions agree very well with each other, and are consistent with the LHCb
results within the still large experimental errors.  For the last two high $q^2$ bins, the LHCb results of $S_3$ ($S_7$) is a little larger (smaller) than our theoretical predictions.

\item[(3)]
For the  CP asymmetry angular coefficients $A_{5,6,8,9}(q^2)$ ,  the PQCD and ``PQCD + Lattice" predictions are very small:  in the range of $10^{-4}$ to $10^{-2}$.
For the LHCb measurements in the six bins, they are clearly  consistent with our theoretical predictions due to still large experimental errors.

\end{enumerate}

In Fig.~\ref{fig:fig4},  we show our theoretical predictions for the $q^2$-dependence of the forward-backward asymmetry $A_{FB}(q^2)$, the optimized observables
$P_{1,2,3}(q^2)$ and $P^{\prime}_{4,5,6,8}(q^2)$ for $B_s \to \phi \mu^+\mu^-$ decay.  Unfortunately, there exist no any experimental measurements
for these observables. We have to wait for future LHCb and Belle-II measurements.
Analogous to Fig.~\ref{fig:fig3}, the vertical grey blocks  in Fig.~\ref{fig:fig4} also  denote the two experimental veto regions of $q^2$: $8.0\! < \!q^2\! <\! 11.0\, GeV^2$ and $12.5\! <\! q^2\! <\! 15.0 \,GeV^2$.

\begin{table}[thb]
\caption{Theoretical predictions for the observables $F^{\phi}_L$, $A_{FB}$, $S_{3,4,7}$, $A_{5,6,8,9}$, $P_{1,2,3}$ and $P^{\prime}_{4,5,6,8}$ integrated over the whole kinematic region
for $B_s \to \phi \ell^+ \ell^-$ decays in the PQCD (the first row) and ``PQCD+Lattice" (the second row) approaches, respectively. }
\label{tab:table7}
\centering
\begin{tabular}{l|ccc|l|ccc}  \hline \hline
${\rm Obs.}$& $\ell=e$& $\ell=\mu$ & $\ell=\tau$ &${\rm Obs.}$& $\ell=e$& $\ell=\mu$ & $\ell=\tau$\\ \hline
$F^{\phi}_L$          & ${0.383}^{+0.003}_{-0.004}$  & ${0.454}^{+0.006}_{-0.007}$ & ${0.396}^{+0.002}_{-0.003}$   &
$-A_{FB}$          & ${0.192}^{+0.003}_{-0.004}$  & ${0.233}^{+0.004}_{-0.004}$ &  ${0.173}^{+0.002}_{-0.002}$ \\
                      & ${0.446}^{+0.012}_{-0.013}$  & ${0.533}^{+0.001}_{-0.001}$  & ${0.442}^{+0.002}_{-0.002}$ &
                  & ${0.152}^{+0.010}_{-0.008}$  & ${0.186}^{+0.005}_{-0.004}$ &  ${0.151}^{+0.002}_{-0.003}$  \\
$-S_3$                 & ${0.120}^{+0.005}_{-0.004}$  & ${0.144}^{+0.005}_{-0.005}$ & ${0.080}^{+0.001}_{-0.001}$&
$-P_1$             & ${0.399}^{+0.014}_{-0.011}$  & ${0.555}^{+0.010}_{-0.012}$ &  ${0.795}^{+0.007}_{-0.004}$ \\
                      & ${0.102}^{+0.009}_{-0.008}$  & ${0.124}^{+0.007}_{-0.006}$ & ${0.075}^{+0.001}_{-0.001}$ &
                  & ${0.381}^{+0.041}_{-0.039}$  & ${0.564}^{+0.026}_{-0.024}$ &  ${0.817}^{+0.007}_{-0.005}$\\
$S_4$                 & ${0.210}^{+0.004}_{-0.004}$  & ${0.258}^{+0.003}_{-0.004}$ & ${0.100}^{+0.001}_{-0.001}$&
$-P_2$             & ${0.213}^{+0.002}_{-0.003}$  & ${0.299}^{+0.001}_{-0.001}$ &  ${0.281}^{+0.003}_{-0.004}$ \\
                      & ${0.201}^{+0.012}_{-0.013}$  & ${0.249}^{+0.007}_{-0.006}$ & ${0.098}^{+0.001}_{-0.001}$ &
                  & ${0.189}^{+0.015}_{-0.013}$  & ${0.282}^{+0.005}_{-0.004}$ &  ${0.268}^{+0.004}_{-0.004}$\\
$10^{3}S_7$           & ${0.350}^{+0.010}_{-0.011}$  & ${0.371}^{+0.012}_{-0.013}$ & ${0.022}^{+0.001}_{-0.000}$&
$10^{2}P_3$       & ${0.103}^{+0.003}_{-0.006}$  & ${0.142}^{+0.002}_{-0.007}$ &  ${0.122}^{+0.003}_{-0.006}$ \\
                      & ${0.392}^{+0.026}_{-0.024}$  & ${0.423}^{+0.040}_{-0.035}$ & ${0.030}^{+0.001}_{-0.000}$&
                  & ${0.118}^{+0.012}_{-0.012}$  & ${0.174}^{+0.007}_{-0.008}$ &  ${0.158}^{+0.002}_{-0.003}$ \\
$-10^{3}A_5$           & ${0.489}^{+0.010}_{-0.009}$  & ${0.585}^{+0.012}_{-0.013}$ & ${0.040}^{+0.000}_{-0.000}$&
$P'_4$            & ${1.049}^{+0.013}_{-0.015}$  & ${1.111}^{+0.011}_{-0.012}$ &  ${1.338}^{+0.002}_{-0.002}$  \\
                      & ${0.518}^{+0.020}_{-0.020}$  & ${0.623}^{+0.041}_{-0.046}$ & ${0.038}^{+0.001}_{-0.001}$ &
                  & ${1.033}^{+0.033}_{-0.037}$  & ${1.098}^{+0.024}_{-0.027}$ &  ${1.345}^{+0.002}_{-0.003}$\\
$-10^{3}A_6$           & ${0.482}^{+0.001}_{-0.001}$  & ${0.577}^{+0.002}_{-0.004}$ & ${0.068}^{+0.000}_{-0.001}$&
$-P'_5$            & ${0.482}^{+0.002}_{-0.004}$  & ${0.524}^{+0.003}_{-0.005}$ &  ${0.427}^{+0.005}_{-0.006}$\\
                      & ${0.429}^{+0.020}_{-0.021}$  & ${0.518}^{+0.038}_{-0.043}$ & ${0.058}^{+0.001}_{-0.001}$ &
                  & ${0.470}^{+0.008}_{-0.010}$  & ${0.515}^{+0.002}_{-0.003}$ &  ${0.408}^{+0.006}_{-0.006}$\\
$10^{4}A_8$           & ${0.541}^{+0.030}_{-0.027}$  & ${0.241}^{+0.017}_{-0.020}$ & ${0.011}^{+0.000}_{-0.000}$&
$-10^{3}P'_6$      & ${0.875}^{+0.031}_{-0.030}$  & ${0.780}^{+0.028}_{-0.029}$ &  ${0.069}^{+0.001}_{-0.001}$\\
                      & ${0.550}^{+0.088}_{-0.095}$  & ${0.237}^{+0.060}_{-0.070}$ & ${0.014}^{+0.000}_{-0.000}$ &
                   & ${1.007}^{+0.089}_{-0.101}$  & ${0.910}^{+0.078}_{-0.089}$ &  ${0.097}^{+0.001}_{-0.001}$ \\
$10^{4}A_9$           & ${0.040}^{+0.002}_{-0.004}$  & ${0.054}^{+0.002}_{-0.004}$ & ${0.013}^{+0.001}_{-0.001}$&
$-10^{2}P'_8$      & ${0.643}^{+0.009}_{-0.003}$  & ${0.640}^{+0.004}_{-0.008}$ &  ${0.287}^{+0.002}_{-0.002}$ \\
                      & ${0.041}^{+0.006}_{-0.008}$  & ${0.056}^{+0.004}_{-0.007}$ & ${0.015}^{+0.001}_{-0.001}$ &
                  & ${0.749}^{+0.004}_{-0.004}$  & ${0.748}^{+0.006}_{-0.006}$ &  ${0.374}^{+0.001}_{-0.001}$\\
 \hline \hline
\end{tabular}
\end{table}

\begin{table}[thb]
\caption{Theoretical predictions for the $q^2$-binned observables ${\cal B}( B_s \to \phi \mu^+ \mu^-)$  (in unit of $10^{-7}$)
in the PQCD (the first low) and ``PQCD+Lattice" (the second row) approach, respectively.  For a comparison, we also list the new LHCb measurements \cite{Aaij:2021pkz}.}
\label{tab:table8}
\centering
\begin{tabular}{l|cc|l|cc}  \hline \hline
$q^2$ bin $(\rm GeV^2)$ & $ {\cal B}(\ell= \mu)$ & LHCb \cite{Aaij:2021pkz}& $q^2$ bin $(\rm GeV^2)$ & $ {\cal B}(\ell= \mu)$ & LHCb \cite{Aaij:2021pkz}\\ \hline
$[0.10,0.98]$     &  ${0.25}^{+0.10}_{-0.07}$ &  ${0.68}\pm{0.06}$ &  $[1.10,2.50]$     &  ${0.19}^{+0.08}_{-0.05}$ &  ${0.44}\pm{0.05}$ \\
                    & ${0.25}^{+0.10}_{-0.07}$  &  $$ & &${0.21}^{+0.07}_{-0.05}$  &  $$ \\
$[2.50,4.00]$     &  ${0.23}^{+0.09}_{-0.07}$ &  ${0.35}\pm{0.04}$ &$[4.00,6.00]$ &  ${0.41}^{+0.18}_{-0.12}$ &  ${0.62}\pm{0.06}$\\
                    &  ${0.27}^{+0.09}_{-0.06}$ &  $$  & & ${0.50}^{+0.14}_{-0.11}$ & $$\\
$[6.00,8.00]$ & ${0.61}^{+0.28}_{-0.19}$  &  ${0.63}\pm{0.06}$ &
$[11.0,12.5]$ & ${0.78}^{+0.38}_{-0.25}$  &  ${0.72}\pm{0.06}$ \\
                    &  ${0.69}^{+0.18}_{-0.13}$ &  $$  &  &${0.76}^{+0.14}_{-0.11}$ &  $$ \\
$[15.00,17.00]$     &  ${1.26}^{+0.64}_{-0.42}$ &  ${1.05}\pm{0.08}$ &
$[17.0,19.0]$ & ${0.86}^{+0.45}_{-0.30}$  &  ${0.84}\pm{0.07}$ \\
                    &  ${1.04}^{+0.11}_{-0.10}$ &  $$  &  &${0.67}^{+0.05}_{-0.06}$ &  $$ \\\hline
$[1.10,6.00]$     &  ${0.83}^{+0.35}_{-0.24}$ &  ${1.41}\pm{0.10}$ &
$[15.0,19.0]$ & ${2.12}^{+1.09}_{-0.71}$  &  ${1.85}\pm{0.13}$ \\
                    &  ${0.98}^{+0.30}_{-0.22}$ &  $$  &  &${1.70}^{+0.17}_{-0.16}$ &                                                  \\
 \hline \hline
\end{tabular}
\end{table}

\begin{table}[thb]
\caption{Theoretical predictions for the $q^2$-binned observables ${\cal B}( B_s \to \phi \ell^+ \ell^-)$  (in unit of $10^{-7}$)   and   $F^\phi_L$  with $\ell=(\mu,\tau)$
in the PQCD (the first low) and ``PQCD+Lattice" (the second row) approach, respectively.
For a comparison, we also list the LHCb measurements as given in Refs.~\cite{Aaij:2015esa,Aaij:2021pkz}. }
\label{tab:table9}
\centering
\begin{tabular}{l|cc|c|cc|c}  \hline \hline
$q^2$ bin $(\rm GeV^2)$ & $ {\cal B}(\ell= \mu)$ & LHCb & ${\cal B}(\ell= \tau)$ & $ F_L^\phi (\ell=\mu)$ & LHCb  \cite{Aaij:2015esa} & $F_L^\phi (\ell=\tau)$ \\ \hline
$[0.1,2.0]$     &  ${0.39}^{+0.16}_{-0.11}$ &  ${1.11}\pm{0.16}$ \cite{Aaij:2015esa} & $-$ &                            ${0.441}^{+0.007}_{-0.008}$ &  ${0.20}\pm{0.09}$ & $-$ \\
                    & ${0.40}^{+0.16}_{-0.11}$  &  $$ & $-$ &                                                   ${0.472}^{+0.011}_{-0.012}$ &  $ $  & $-$ \\
$[2.0,5.0]$     &  ${0.48}^{+0.20}_{-0.14}$ &  ${0.77}\pm{0.14}$ \cite{Aaij:2015esa}   & $-$&                              ${0.738}^{+0.008}_{-0.009}$ &  ${0.68}\pm{0.15}$ & $-$\\
                    & ${0.58}^{+0.18}_{-0.13}$  &  $$ & $-$ &                                                 ${0.796}^{+0.007}_{-0.007}$ &  $ $ & $-$ \\
$[5.0,8.0]$     &  ${0.84}^{+0.37}_{-0.25}$ &  ${0.96}\pm{0.15}$  \cite{Aaij:2015esa} & $-$&                                        ${0.584}^{+0.008}_{-0.009}$ &  ${0.54}\pm{0.10}$ & $-$\\
                    &  ${0.96}^{+0.25}_{-0.19}$ &  $$ & $-$ &                                                  ${0.682}^{+0.010}_{-0.008}$ &  $ $ & $  -$ \\
$[11.0,12.5]$ &  ${0.78}^{+0.38}_{-0.25}$ &  ${0.71}\pm{0.12}$ \cite{Aaij:2015esa}  & $-$&                                  $        {0.433}^{+0.003}_{-0.005}$ &  ${0.29}\pm{0.11}$ & $-$\\
                    &  ${0.76}^{+0.14}_{-0.11}$ &  $0.72\pm0.06$\cite{Aaij:2021pkz} & $-$ &                                                  ${0.524}^{+0.008}_{-0.008}$ &  $ $ & $- $ \\
$[15.0,17.0]$ & ${1.12}^{+0.61}_{-0.40}$  &  ${0.90}\pm{0.13}$ \cite{Aaij:2015esa}  & ${0.45}^{+0.22}_{-0.14}$&     $                {0.368}^{+0.001}_{-0.001}$ &  ${0.23}\pm{0.09}$  & ${0.421}^{+0.003}_{-0.004}$\\
                    & ${0.99}^{+0.11}_{-0.10}$  &  $1.05\pm0.08$\cite{Aaij:2021pkz} & ${0.39}^{+0.04}_{-0.04}$&                             ${0.412}^{+0.004}_{-0.004}$ &  $$              & ${0.479}^{+0.003}_{-0.003}$ \\
$[17.0,19.0]$ & ${0.80}^{+0.41}_{-0.27}$  &  ${0.75}\pm{0.13}$  \cite{Aaij:2015esa} & ${0.36}^{+0.18}_{-0.12}$&       $              {0.346}^{+0.001}_{-0.001}$ &  ${0.40}\pm{0.14}$  & ${0.365}^{+0.001}_{-0.002}$\\
                    &  ${0.67}^{+0.05}_{-0.05}$ &  $0.82\pm0.07$\cite{Aaij:2021pkz} & ${0.29}^{+0.02}_{-0.02}$ &                      ${0.363}^{+0.001}_{-0.001}$ &  $$              & ${0.391}^{+0.001}_{-0.001}$ \\ \hline
$[1.0,6.0]$     &  ${0.93}^{+0.29}_{-0.27}$ &  ${1.29}\pm{0.19}$  \cite{Aaij:2015esa} & $-$&                                        ${0.708}^{+0.007}_{-0.009}$ &  ${0.63}\pm{0.09}$ & $-$\\
                    &  ${1.10}^{+0.34}_{-0.25}$ &  $$ & $-$ &                                                  ${0.777}^{+0.008}_{-0.006}$ &  $$ & $-$ \\
$[15.0,19.0]$ & ${1.99}^{+1.02}_{-0.71}$  &  ${1.62}\pm{0.20}$ \cite{Aaij:2015esa}  & ${0.82}^{+0.40}_{-0.26}$&         $            {0.359}^{+0.002}_{-0.001}$ &  ${0.29}\pm{0.07}$  & ${0.396}^{+0.003}_{-0.003}$\\
                    &  ${1.60}^{+0.16}_{-0.16}$ &  $1.85\pm0.13$\cite{Aaij:2021pkz} & ${0.68}^{+0.06}_{-0.06}$ &                            ${0.394}^{+0.003}_{-0.003}$ &                & ${0.442}^{+0.002}_{-0.001}$ \\
 \hline \hline
\end{tabular}
\end{table}

In Table \ref{tab:table7}, we list the theoretical predictions for the values of the observables $F^{\phi}_L$, $A_{FB}$, $S_{3,4,7}$, $A_{5,6,8,9}$, $P_{1,2,3}$ and $P^{\prime}_{4,5,6,8}$, obtained  after the integrations
over the whole kinematic region of $q^2$ for the semileptonic decays $B_s \to \phi \ell^+ \ell^-$ with $\ell=(e,\mu,\tau)$  in the PQCD (the first row) and ``PQCD+Lattice" (the second row) approaches, respectively.
Of course,  the regions corresponding to resonance $J/\psi(1S)$ and $\psi(2S)$, say $8.0\! < \!q^2\! <\! 11.0\, GeV^2$ and $12.5\! <\! q^2\! <\! 15.0 \,GeV^2$ numerically,  are also cut off here.
The total errors are the combinations of the individual errors from the form factors, the renormalization scales and the relevant CKM matrix elements.
The above theoretical predictions should  be tested in the near future LHCb and Belle-II experiments.
For the considered $B_s$ meson decays, one should consider the effects from the $B_s$-$\bar B_s$ mixing \cite{Hiller2015}.
The theoretical framework for examining the time-dependent decays with the inclusion of such mixing effects can been found in Ref.~\cite{Descotes-Genon:2015hea}.
The authors of Ref.~\cite{Descotes-Genon:2015hea} proved that the mixing effects on the values of decay rates and CP  averaged observables are generally within a few percent
and could be neglected.

\subsection{The \texorpdfstring{$q^2$}{}-binned predictions}

\begin{table}[thb]
\caption{Theoretical predictions for the $q^2$-binned observables $S_{3,4,7}$ of the decays $B_s \to \phi \mu^+ \mu^-$
in the PQCD (the first low) and ``PQCD+Lattice" (the second row) approaches. For a comparison, we also list the LHCb measured values \cite{Aaij:2015esa}.}
\label{tab:table10}
\centering
\begin{tabular}{l|cc|cc|cc}  \hline \hline
                                          &\multicolumn{2}{c|}{$S_3$}& \multicolumn{2}{c|}{$S_4$}  & \multicolumn{2}{c}{$S_7$}  \\  \cline{2-7}
$q^2$ bin $ (\rm GeV^2)$ & Theor.           & LHCb                              &  Theor.                     & LHCb                                        &  Theor. ($10^{-3}$)                         & LHCb \\  \hline
$[0.1,2.0]$    & ${0.003}^{+0.000}_{-0.000}$  &  ${-0.05}\pm{0.13}$         & ${-0.054}^{+0.001}_{-0.001}$ &  ${-0.27}\pm{0.23}$                   & $1.571^{+0.001}_{-0.002}$ &  ${-0.04}\pm{0.12}$  \\
                   & ${0.002}^{+0.000}_{-0.001}$  &                                       & ${-0.053}^{+0.000}_{-0.000}$ &                                                  & $1.551^{+0.003}_{-0.006}$ &                                 \\
$[2.0,5.0]$    & ${-0.024}^{+0.002}_{-0.002}$ &  ${-0.06}\pm{0.21}$         & ${0.191}^{+0.004}_{-0.004}$ &  ${0.47}\pm{0.37}$                       & $1.065^{+0.003}_{-0.003}$ &  ${0.03}\pm{0.21}$      \\
                   & ${-0.021}^{+0.001}_{-0.001}$ &                                       & ${0.177}^{+0.003}_{-0.003}$ &                                                   & ${0.979}^{+0.017}_{-0.018}$ &                                     \\
$[5.0,8.0]$    & ${-0.057}^{+0.003}_{-0.003}$ &  ${-0.10}\pm{0.25}$         & ${0.270}^{+0.002}_{-0.002}$ &  ${0.10}\pm{0.17}$                        & ${0.453}^{+0.003}_{-0.003}$ &  ${-0.04}\pm{0.18}$   \\
                   & ${-0.050}^{+0.004}_{-0.003}$ &                                        & ${0.259}^{+0.003}_{-0.003}$ &                                                   & ${0.453}^{+0.004}_{-0.005}$ &                                  \\
$[11.0,12.5]$ & ${-0.124}^{+0.004}_{-0.003}$ &  ${-0.19}\pm{0.21}$         & ${0.296}^{+0.001}_{-0.001}$ &  ${0.47}\pm{0.25}$                        & ${0.153}^{+0.002}_{-0.002}$ &  ${0.00}\pm{0.16}$   \\
                    & ${-0.115}^{+0.002}_{-0.001}$ &                                       & ${0.303}^{+0.001}_{-0.001}$ &                                                   & ${0.185}^{+0.001}_{-0.001}$ &                                \\
$[15.0,17.0]$ & ${-0.219}^{+0.003}_{-0.003}$ &  ${-0.06}\pm{0.18}$         & ${0.314}^{+0.001}_{-0.001}$ &  ${0.03}\pm{0.15}$                       & ${0.052}^{+0.001}_{-0.001}$ &  ${-0.12}\pm{0.15}$  \\
                    & ${-0.213}^{+0.001}_{-0.001}$ &                                       & ${0.323}^{+0.001}_{-0.001}$ &                                                   & ${0.071}^{+0.001}_{-0.001}$ &                                  \\
$[17.0,19.0]$ & ${-0.283}^{+0.001}_{-0.001}$ &  ${-0.07}\pm{0.25}$         & ${0.325}^{+0.001}_{-0.001}$ &  ${0.39}\pm{0.30}$                        & ${0.019}^{+0.001}_{-0.001}$ &  ${-0.20}\pm{0.26}$   \\
                    & ${-0.281}^{+0.001}_{-0.001}$ &                                       & ${0.329}^{+0.001}_{-0.001}$ &                                                   & ${0.027}^{+0.001}_{-0.001}$ &                                    \\ \hline
$[1.0,6.0]$     & ${-0.026}^{+0.002}_{-0.002}$ &  ${-0.02}\pm{0.13}$         & ${0.180}^{+0.004}_{-0.004}$ &  ${0.19}\pm{0.14}$                       & ${1.063}^{+0.001}_{-0.001}$ &  ${0.03}\pm{0.14}$ \\
                    & ${-0.023}^{+0.001}_{-0.001}$ &                                       & ${0.169}^{+0.001}_{-0.001}$ &                                                   & ${0.985}^{+0.020}_{-0.022}$ &                                    \\
$[15.0,19.0]$ & ${-0.245}^{+0.002}_{-0.002}$ &  ${-0.09}\pm{0.12}$          & ${0.318}^{+0.001}_{-0.001}$ &  ${0.14}\pm{0.11}$                       & ${0.038}^{+0.002}_{-0.002}$ &  ${-0.13}\pm{0.11}$        \\
                    & ${-0.239}^{+0.001}_{-0.001}$ &                                        & ${0.325}^{+0.001}_{-0.001}$ &                                                   & ${0.054}^{+0.001}_{-0.001}$ &                               \\
 \hline\hline
\end{tabular}
\end{table}

For $B_s \to \phi \mu^+ \mu^-$ decay mode, the LHCb Collaboration has reported their experimental measurements for many physical observables in several $q^2$ bins  \cite{Aaij:2015esa,Aaij:2021pkz}.
In order to compare our theoretical predictions with  the LHCb results bin by bin, we  make the same choices of the $q^2$ bins  as LHCb did, calculate and show our theoretical
predictions for the branching ratio ${\cal B}(B_s \to \phi \ell^+ \ell^-)$  and the asymmetry $F_L^{\phi}$ with $\ell =(\mu,\tau)$ in Table \ref{tab:table8} and \ref{tab:table9},
and the observables $S_{3,4,7}$ with $\ell=\mu$ in Table \ref{tab:table10}.
For observables $S_7$ and $A_{5,6,8,9}$,  in fact,  our theoretical predictions for their values are very small,
say in the range of $10^{-3}-10^{-4}$ in magnitude, but still agree with the LHCb measurements in different bins \cite{Aaij:2015esa} due to still large experimental errors.
For the observables $P_3$ and $P^\prime_{6,8}$, they are also very small in size:  in the range of  $10^{-3}-10^{-4}$  and there exist no corresponding data at present.
For observables $P_{1,2}$ and $P^\prime_{4,5}$,  on the other hand, although there exist no experimental measurements for them at present,
they are relatively large in size and may be measured in the near future LHCb and Belle-II experiments,  so we calculate and list the theoretical predictions
of these observables bin by bin for the cases of $\ell=(\mu,\tau)$ in Table \ref{tab:table11} and \ref{tab:table12}.
Very recently, LHCb reported some new measurements for the angular observables of $B_s \to \phi \mu^+\mu^-$ decay \cite{LHCb-2107}  in the $q^2$ bins different from those in
their previous work \cite{Aaij:2015esa}, which will be studied in our next work.

The definitions of the $q^2$-binned observables  are the following:
\beq
{\cal B} (q^2_1,q^2_2)  =  \int^{q^2_2}_{q^2_1}dq^2 \frac{d\calb(B_s\to \phi \ell^+\ell^-)}{dq^2} , \label{eq:brq1q2}
\eeq
\beq
F_L^{\phi} (q^2_1,q^2_2)= \frac{\int^{q^2_2}_{q^2_1}dq^2 [3(I^c_{1}+\bar{I}^c_{1})-(I^c_{2}+\bar{I}^c_{2})]}{4\int^{q^2_2}_{q^2_1}dq^2 [d(\Gamma+\bar{\Gamma})/dq^2]}, \quad \label{eq:flphi01}
\eeq
\beq
{A}_{\rm FB}(q^2_1,q^2_2) = \frac{3\int^{q^2_2}_{q^2_1}dq^2 (I^s_{6}+\bar{I}^s_{6})}{4\int^{q^2_2}_{q^2_1}dq^2[d(\Gamma+\bar{\Gamma})/dq^2]}, \label{eq:afbq112}
\eeq
\beq
S_{3,4,7}(q^2_1,q^2_2) = \frac{\int_{q_{1}^{2} }^{q_{2}^{2}}dq^{2}(I_{3,4,7}+\bar{I}_{3,4,7})}{\int_{q_{1}^{2} }^{q_{2}^{2}}dq^{2}[d(\Gamma +\bar{\Gamma})/dq^{2}]} , \quad \label{eq:s347}
\eeq
\beq
 P_1 (q^2_1,q^2_2) &=& \frac{ \int^{q^2_2}_{q^2_1}dq^2 (S_{3})}{2 \int^{q^2_2}_{q^2_1}dq^2 (S^s_{2})}, \quad
 P_2 (q^2_1,q^2_2) =\frac{\int^{q^2_2}_{q^2_1}dq^2 (\beta_\ell S^s_{6})}{8 \int^{q^2_2}_{q^2_1}dq^2 (S^s_{2})}, \label{eq:p1p2}
 \eeq
 \beq
 P^\prime_4 (q^2_1,q^2_2) &=&\frac{\int^{q^2_2}_{q^2_1}dq^2 (S_4)}{\sqrt{-\int^{q^2_2}_{q^2_1}dq^2 (S^c_2 S^s_2)}} , \quad
 P^\prime_5(q^2_1,q^2_2) =\frac{\int^{q^2_2}_{q^2_1}dq^2 (\beta_\ell S_5)}{2\sqrt{-\int^{q^2_2}_{q^2_1}dq^2 (S^c_2 S^s_2)}} \label{eq:p4pp5p}.
\eeq

\begin{table}[thb]
\caption{Theoretical predictions for the $q^2$-binned observables $ A^\mu_{FB}$, $ P_{1,2}$  of the decays $B_s \to \phi \ell^+ \ell^-$ with $\ell=(\mu,\tau)$
in the PQCD (the first low) and ``PQCD+Lattice" (the second row) approaches.}
\label{tab:table11}
\centering
\begin{tabular}{l|cc|cc|cc}  \hline \hline
$q^2$ bin $(\rm GeV^2)$& $ A^\mu_{FB}$& $ A^\tau_{FB}$ & $ P_1(\ell=\mu)$& $ P_1(\ell=\tau)$ &  $ P_2(\ell=\mu)$& $ P_2(\ell=\tau)$  \\ \hline
$[0.1,2.0]$       & ${0.131}^{+0.003}_{-0.003}$ &   $-$                                                & ${0.015}^{+0.001}_{-0.001}$ &   $-$                           & ${0.206}^{+0.001}_{-0.001}$ &   $-$\\
                     & ${0.122}^{+0.003}_{-0.002}$ &   $-$                                                    & ${0.013}^{+0.001}_{-0.001}$ &  $-$                            & ${0.204}^{+0.001}_{-0.001}$ &   $-$ \\
$[2.0,5.0]$        & ${-0.047}^{+0.004}_{-0.003}$ &   $-$                                             & ${-0.197}^{+0.007}_{-0.007}$ &   $-$                           & ${-0.128}^{+0.003}_{-0.003}$ &   $-$\\
                     & ${-0.038}^{+0.001}_{-0.002}$ &   $-$                                                  & ${-0.225}^{+0.005}_{-0.006}$ &   $-$                            & ${-0.132}^{+0.002}_{-0.002}$ &   $-$ \\
$[5.0,8.0]$        & ${-0.265}^{+0.006}_{-0.005}$ &   $-$                                             & ${-0.280}^{+0.009}_{-0.008}$ &   $-$                           & ${-0.431}^{+0.001}_{-0.001}$ &   $-$\\
                     & ${-0.200}^{+0.006}_{-0.006}$ &   $-$                                                  & ${-0.323}^{+0.003}_{-0.004}$ &   $-$                             & ${-0.427}^{+0.001}_{-0.001}$ &   $-$ \\
$[11.0,12.5]$   & ${-0.365}^{+0.001}_{-0.001}$ &   $-$                                                & ${-0.440}^{+0.010}_{-0.008}$ &   $-$                              & ${-0.431}^{+0.002}_{-0.003}$ &   $-$\\
                     & ${-0.298}^{+0.005}_{-0.005}$ &   $-$                                                & ${-0.489}^{+0.002}_{-0.003}$ &   $-$                                 & ${-0.420}^{+0.001}_{-0.001}$ &   $-$ \\
$[15.0,17.0]$   & ${-0.326}^{+0.001}_{-0.001}$ &   ${-0.188}^{+0.001}_{-0.001}$   & ${-0.697}^{+0.009}_{-0.006}$ &   ${-0.707}^{+0.009}_{-0.005}$        & ${-0.345}^{+0.003}_{-0.004}$ &   ${-0.341}^{+0.003}_{-0.004}$\\
                   & ${-0.290}^{+0.005}_{-0.005}$ &   ${-0.160}^{+0.003}_{-0.003}$          & ${-0.730}^{+0.007}_{-0.005}$ &   ${-0.738}^{+0.006}_{-0.006}$     & ${-0.330}^{+0.003}_{-0.003}$ &   ${-0.326}^{+0.003}_{-0.004}$ \\
$[17.0,19.0]$   & ${-0.226}^{+0.002}_{-0.003}$ &   ${-0.153}^{+0.001}_{-0.002}$    & ${-0.869}^{+0.005}_{-0.003}$ &   ${-0.875}^{+0.005}_{-0.003}$    & ${-0.231}^{+0.002}_{-0.003}$ &   ${-0.225}^{+0.003}_{-0.004}$\\
                   & ${-0.208}^{+0.004}_{-0.003}$ &   ${-0.137}^{+0.003}_{-0.002}$           & ${-0.884}^{+0.004}_{-0.004}$ &   ${-0.890}^{+0.005}_{-0.003}$  & ${-0.219}^{+0.003}_{-0.004}$ &   ${-0.213}^{+0.004}_{-0.004}$ \\ \hline
\hline
\end{tabular}
\end{table}

\begin{table}[thb]
\caption{Theoretical predictions for the $q^2$-binned optimized observables  $ P'_4$  and  $ P'_5$  of the decays $B_s \to \phi \ell^+ \ell^-$ with $\ell=(e,\mu,\tau)$
in the PQCD (the first low) and ``PQCD+Lattice" (the second row) approach. }
\label{tab:table12}  
\centering
\begin{tabular}{l|ccc|ccc}  \hline \hline
$q^2$ bin $(\rm GeV^2)$& $ P'_4(\ell=e)$ & $ P'_4(\ell=\mu)$ & $ P'_4(\ell=\tau)$ & $ P'_5(\ell=e)$ & $ P'_5(\ell=\mu)$ & $ P'_5(\ell=\tau)$  \\ \hline
$[0.1,2.0]$    & ${-0.322}^{+0.005}_{-0.005}$  & ${-0.289}^{+0.005}_{-0.005}$ &   $-$                                     & ${0.484}^{+0.001}_{-0.002}$  & ${0.459}^{+0.001}_{-0.002}$ &   $-$\\
                 & ${-0.318}^{+0.001}_{-0.001}$  & ${-0.285}^{+0.001}_{-0.001}$ &   $-$                                         & ${0.475}^{+0.002}_{-0.001}$  & ${0.449}^{+0.001}_{-0.001}$ &   $-$ \\
$[2.0,5.0]$    & ${0.903}^{+0.007}_{-0.007}$  & ${0.903}^{+0.007}_{-0.007}$ &   $-$                                       & ${-0.606}^{+0.001}_{-0.002}$  & ${-0.607}^{+0.001}_{-0.002}$ &   $-$\\
                 & ${0.916}^{+0.001}_{-0.001}$  & ${0.916}^{+0.001}_{-0.001}$ &   $-$                                          & ${-0.608}^{+0.004}_{-0.003}$  & ${-0.608}^{+0.004}_{-0.004}$ &   $-$ \\
$[5.0,8.0]$    & ${1.110}^{+0.003}_{-0.005}$  & ${1.110}^{+0.004}_{-0.005}$ &   $-$                                      & ${-0.816}^{+0.005}_{-0.006}$  & ${-0.816}^{+0.005}_{-0.006}$ &   $-$\\
                 & ${1.127}^{+0.002}_{-0.001}$  & ${1.127}^{+0.002}_{-0.001}$ &   $-$                                           & ${-0.796}^{+0.003}_{-0.001}$  & ${-0.796}^{+0.002}_{-0.002}$ &   $-$ \\
$[11.0,12.5]$  & ${1.198}^{+0.002}_{-0.005}$  & ${1.198}^{+0.002}_{-0.005}$ &   $-$                                       & ${-0.727}^{+0.005}_{-0.007}$  & ${-0.727}^{+0.005}_{-0.007}$ &   $-$\\
                 & ${1.217}^{+0.001}_{-0.001}$  & ${1.217}^{+0.001}_{-0.001}$ &   $-$                                           & ${-0.698}^{+0.001}_{-0.001}$  & ${-0.698}^{+0.002}_{-0.001}$ &   $-$ \\
$[15.0,17.0]$  & ${1.302}^{+0.002}_{-0.004}$  & ${1.302}^{+0.002}_{-0.004}$ &   ${1.306}^{+0.002}_{-0.004}$ & ${-0.533}^{+0.005}_{-0.007}$  & ${-0.533}^{+0.005}_{-0.007}$ &   ${-0.525}^{+0.006}_{-0.008}$\\
                 & ${1.314}^{+0.002}_{-0.003}$  & ${1.314}^{+0.002}_{-0.003}$ &   ${1.317}^{+0.003}_{-0.002}$        & ${-0.506}^{+0.005}_{-0.006}$  & ${-0.506}^{+0.005}_{-0.006}$ &   ${-0.498}^{+0.006}_{-0.006}$ \\
$[17.0,19.0]$  & ${1.366}^{+0.001}_{-0.002}$  & ${1.366}^{+0.001}_{-0.002}$ &   ${1.369}^{+0.001}_{-0.002}$ & ${-0.341}^{+0.004}_{-0.006}$  & ${-0.341}^{+0.004}_{-0.006}$ &   ${-0.331}^{+0.004}_{-0.006}$\\
                 & ${1.371}^{+0.001}_{-0.001}$  & ${1.371}^{+0.001}_{-0.001}$ &   ${1.374}^{+0.001}_{-0.002}$    & ${-0.323}^{+0.005}_{-0.006}$  & ${-0.323}^{+0.005}_{-0.006}$ &   ${-0.314}^{+0.006}_{-0.006}$ \\ \hline
\hline
\end{tabular}
\end{table}


From the numerical values as shown in Fig.~\ref{fig:fig3} and in Table \ref{tab:table8}, \ref{tab:table9} and \ref{tab:table10}, we find the following points about the relevant physical
observables of  the considered $B_s \to \phi  \mu^+ \mu^- $ decays in bins:
\begin{enumerate}
\item[(1)]
For  $B_s \to \phi \mu^+ \mu^-$ decay, besides the good consistency between the theory and the LHCb data for the integrated total branching ratio ${\cal B}(B_s \to \phi \mu^+ \mu^-)$ as listed in
Eq.~(\ref{eq:brtot1}),  the  PQCD and ``PQCD+Lattice"   predictions for  $\calb(B_s \to \phi \mu^+ \mu^-)$ in most bins do agree well with the measured ones within $2\sigma$ errors.
For the first low-$q^2$ bin $0.1 < q^2 < 2$ $({\rm GeV^2})$,  however,  the central value of the LHCb result $1.11\pm 0.16$  is larger than the theoretical ones by roughly a factor of three.
The LHCb results of  ${\cal B}(B_s \to \phi \mu^+ \mu^-)$ in different bins of $q^2$ as listed in the third column of Table \ref{tab:table9}
are obtained from the results as given in  Refs.~\cite{Aaij:2015esa,Aaij:2021pkz}
by multiplying  the LHCb measured values of differential decay rate $d{\cal B}(B_s \to \phi \mu^+\mu^-)/dq^2$ with the width of the corresponding bin $(q_2^2-q_1^2) $.
The theoretical errors of our theoretical predictions of the branching ratios in bins are still relatively large,  while the differences between the PQCD and ``PQCD+Lattice"   predictions
for  ${\cal B}(B_s \to \phi \ell^+ \ell^-)$ with $\ell=(\mu,\tau)$ are small.

\item[(2)]
In the first low-$q^2$ bin  $0.1 < q^2 < 2$ $({\rm GeV^2})$,   both the PQCD  and "PQCD+Lattice"  predictions for $F_L^\phi(\ell=\mu)$  are larger than the LHCb measured results
$F_L^\phi(\ell=\mu)|_{{\rm LHCb}}=0.20^{+0.08}_{-0.09} \pm 0.02$  \cite{Aaij:2015esa}.
For other bins,  both PQCD and ``PQCD+Lattice"  predictions of $F^\phi_{L}$  for  muon mode do  agree very well  with currently  available  LHCb measured values \cite{Aaij:2015esa}
within $2\sigma$ errors.  It is worth of remaining that our theoretical predictions of $F_L^\phi$  have a little error of  $\sim 2\%$ due to the strong cancellation of the theoretical  errors in the ratios.
The theoretical predictions for ${\cal B}(B_s \to \phi \tau^+ \tau^-)$ and $F_L^\phi (\ell=\tau)$ in different bins of $q^2$ as listed in Table \ref{tab:table9} will be tested
by future experimental measurements,

\item[(3)]
For the observables $S_{3,4,7}$,  as listed in Table \ref{tab:table10},  the PQCD and  ``PQCD+Lattice" predictions for their values in all bins are in the range
of $10^{-3} - 10^{-1}$, and show  a good agreement with the LHCb measured values \cite{Aaij:2015esa} .
The  errors of the theoretical predictions are also very small, $\sim 2\%$ in magnitude,  because of their nature of the ratios.
In all bins, the LHCb measured values of $S_{3,4,7}$  are still consistent with zero due to their still large errors, which is  a clear feature as can be seen easily
from the numerical values in Table \ref{tab:table10} and the crosses in Fig.~\ref{fig:fig3}.

\end{enumerate}

In  Table \ref{tab:table11} and \ref{tab:table12},  we show the PQCD and ``PQCD+Lattice" predictions for the physical observables $A_{FB}^{\mu,\tau}, P_{1,2}^{\mu,\tau}$ and
$P^{\prime}_{4,5}(\ell=e,\mu,\tau)$  in  six bins.  These physical observables could be tested in the near future LHCb and Belle-II experiments.


\section{Summary} \label{sec:6}

In this paper,  we made a systematic study of the semileptonic decays $B_s \to \phi \ell^+ \ell^-$ with $\ell^-=(e^-,\mu^-,\tau^-)$  using the PQCD
and the  ``PQCD+Lattice" factorization approach respectively.
We first evaluated all relevant form factors  in the low $q^2$ region using the PQCD approach, and we  also took currently available lattice QCD results
at the high-$q^2$ points  $q^2=(12, 16,18.9)$ GeV$^2$ as additional input to improve  the extrapolation  of  the form factors  from the low to the high-$q^2$ region.
We calculated the branching ratios ${\cal B}( B_s \to \phi \ell^+ \ell^-)$,  the CP averaged $\phi$ longitudinal polarization fraction $F_L(q^2)$, the forward-backward asymmetry
$\cala_{FB}(q^2)$,  the CP averaged angular coefficients $S_{3,4,7}(q^2)$, the CP asymmetry angular coefficients $A_{5,6,8,9}(q^2)$,  the optimized observables $P_{1,2,3}(q^2)$
and $P^\prime_{4,5,6,8}(q^2)$.  For $B_s \to \phi \mu^+\mu^-$ decay mode,  we  calculated the binned values of the branching ratio ${\cal B}( B_s \to \phi \mu^+ \mu^-)$,
the observables $F_L^\phi$ and $S_{3,4,7}$ in the same bins as defined by LHCb Collaboration \cite{Aaij:2015esa} in order to compare our theoretical predictions with those currently available
LHCb measurements bin by bin directly.

Based on the analytical evaluations, the numerical results and the phenomenological analysis,  we found the following main points:
\begin{enumerate}
\item[(1)]
For the branching ratio ${\cal B}(B_s \to \phi \mu^+ \mu^-)$,  the  PQCD and ``PQCD+Lattice" prediction are $(7.07^{+3.43}_{-2.34})\times 10^{-7}$ and  $(6.76^{+1.52}_{-1.25})\times 10^{-7}$  respectively,
which agree well with the LHCb measured values \cite{Aaij:2015esa,Aaij:2021pkz}  and the QCDSR prediction within still large errors.
For the electron and tau mode, our theoretical predictions for their decay rates are also well consistent with the corresponding QCDSR predictions and to be tested by future experimental measurements.

\item[(2)]
For the ratios of the branching ratios $R_\phi^{e\mu}$ and  $R_\phi^{\mu\tau}$,  the PQCD and ``PQCD+Lattice" predictions agree with each other and with small theoretical
errors because of the strong cancellation of the theoretical errors in such ratios. We suggest the LHCb and Belle-II collaboration to measure these ratios.

\item[(3)]
For the longitudinal polarization $F_{L}$,  both PQCD and "PQCD+Lattice" predictions agree  with the LHCb measurements in the considered bins within the errors.
For  the CP averaged angular coefficients $S_{3,4,7}$, the PQCD and "PQCD+Lattice" predictions in all bins are small in magnitude,  in the range
of $10^{-3} - 10^{-1}$,  and agree well with the LHCb results within the still large experimental errors.
For the  CP asymmetry angular coefficients $A_{5,6,8,9}$ ,  the PQCD and "PQCD+Lattice" predictions are very small, in the range of $10^{-4}-10^{-2}$, and
clearly consistent with the LHCb measurements in the six bins.

\item[(4)]
For the physical observables $A_{FB}^{l}$, $P_{1,2,3}$ and $P^{\prime}_{4,5,6,8}$,   the experimental measurements are still absent now,
we think that the PQCD and ``PQCD+Lattice" predictions for these physical observables  will be tested in the near future LHCb and Belle-II experiments.
\end{enumerate}

\begin{acknowledgments}

This work was supported by the National Natural Science Foundation of China under Grant  No.~11775117 and 11235005.

\end{acknowledgments}


\end{document}